\documentclass[aps,
               pra,
               twocolumn,
               amsmath,
               amssymb,
               superscriptaddress,
               reprint,
               10pt
              ]{revtex4-2}

\usepackage{url}
\usepackage{txfonts}

\usepackage{amsmath, amssymb}

\usepackage{pstricks}
\usepackage{graphicx}
\usepackage{tikz}
\usepackage[T1]{fontenc}
\usepackage{color}
\usepackage[colorlinks=true,breaklinks=true,allcolors=blue]{hyperref}

\usepackage{tabularx}
\usepackage{multirow}
\usepackage{array}
\newcolumntype{L}[1]{>{\raggedright\let\newline\\\arraybackslash\hspace{0pt}}m{#1}}
\newcolumntype{C}[1]{>{\centering\let\newline\\\arraybackslash\hspace{0pt}}m{#1}}
\newcolumntype{R}[1]{>{\raggedleft\let\newline\\\arraybackslash\hspace{0pt}}m{#1}}

\usepackage{physics}
\usepackage{xparse}
\usepackage[version=4]{mhchem}
\usepackage{lipsum}
\usepackage[normalem]{ulem}

\renewcommand{\i}{\mathrm{i}}
\newcommand{\e}{\mathrm{e}}

\newcommand{\kvperpq}{\vb{k}}
\newcommand{\kperpq}{k}

\newcommand{\eq}[1]{(\ref{eq:#1})}
\newcommand{\Eq}[1]{Eq.\,\eqref{eq:#1}}

\newcommand{\Fig}[1]{Fig.~\ref{fig:#1}}
\newcommand{\fig}[1]{\ref{fig:#1}}

\newcommand{\Sect}[1]{Sect.~\ref{sec:#1}}
\newcommand{\Subsect}[1]{Sect.~\ref{subsec:#1}}

\newcommand{\App}[1]{App.~\ref{app:#1}}

\renewcommand*\exp[1]{\mathrm{exp} \qty( #1 )}

\newcommand*\aho{a_\mathrm{ho}}
\newcommand*\add{a_\mathrm{dd}}

\newcommand*\epsdd{\epsilon_\mathrm{dd}}
\newcommand*\xih{\xi_\mathrm{h}}

\renewcommand*\i{\mathrm{i}}
\newcommand*\Cdd{C_\mathrm{dd}}

\newcommand{\sludilute}{\emph{ultradilute}\ }
\newcommand{\slquantum}{\emph{quantum}\ }
\newcommand{\udilute}{{ultradilute}\ }
\newcommand{\quantum}{{quantum}\ }

\makeatletter
\let\cat@comma@active\@empty
\makeatother

\begin{document}

\title{Anomalous non-thermal fixed point in a quasi-two-dimensional dipolar Bose gas}

\author{Niklas Rasch}
\email{niklas.rasch@kip.uni-heidelberg.de}
\affiliation{Kirchhoff-Institut f\"ur Physik, 
	Ruprecht-Karls-Universit\"at Heidelberg,
	Im Neuenheimer Feld 227, 
	69120 Heidelberg, Germany}

\author{Lauriane Chomaz}
\affiliation{Physikalisches Institut,
	Ruprecht-Karls-Universit\"at Heidelberg,
	Im Neuenheimer Feld 226,
	69120 Heidelberg, Germany}

\author{Thomas Gasenzer}
\email{t.gasenzer@uni-heidelberg.de}
\affiliation{Kirchhoff-Institut f\"ur Physik, 
             Ruprecht-Karls-Universit\"at Heidelberg,
             Im Neuenheimer Feld 227, 
             69120 Heidelberg, Germany}
\affiliation{Institut f\"ur Theoretische Physik, 
			Ruprecht-Karls-Universit\"at Heidelberg,
			Philosophenweg 16, 69120 Heidelberg, Germany}

\date{\today}

\begin{abstract}
	The emergence of distinctly sub-diffusive scaling in the vicinity of an anomalous non-thermal fixed point is discussed in a quasi-two-dimensional dipolar Bose gas in the superfluid phase, carrying ensembles of vortices and antivortices with zero net angular momentum.
	The observed scaling behavior reflects coarsening dynamics driven by the mutual annihilation of vortices and antivortices, with the mean inter-defect distance growing algebraically over time as $\ell_\text{v}(t)\sim t^{\,\beta}$.
	A sub-diffusive ($\beta<1/2$) exponent $\beta\approx0.2$ is extracted for various parameter regimes, initial conditions, and dipolar configurations from both scaling occupation-number spectra and the evolution of inter-defect distances as well as the corresponding total vortex densities.
	As vortex-antivortex annihilation progresses, excitations of the background condensate increase. This gives rise to a transition in the scaling behavior at late times, toward a non-thermal fixed point governed by diffusion-type scaling with $\beta\approx1/2$ as expected for the mutual annihilation of well-separated vortex-antivortex dipoles.
	While the temporal scaling with $\beta$ does not depend significantly on the strength and anisotropy of the dipolar interactions and thus underlines the universality of the anomalous as well as diffusion-type non-thermal fixed points, we find distinctly different vortex patterns resulting in the dipolar case. 
	While in the superfluid with contact interactions only, same-sign vortices tend to cluster and form large-scale eddies, in the dipolar and tilted cases, roton excitations appear to prevent such motion, giving rather rise to a maximisation of distances between vortices of either sign.
\end{abstract}

\maketitle

\section{Introduction}
\label{sec:introduction}

Scaling phenomena have long been central to physics, with early studies focusing on fluid turbulence \cite{Frisch1995a}.
On microscopic scales, where quantum statistics and coherence prevail, analogous scaling phenomena emerge in low-energy as well as relativistic systems, including quantum turbulence in superfluids 
\cite{Svistunov2001a,
Volovik2004a,
Vinen2006a,
Tsubota2008a,
Proment2010a,
Micha2003a,
Henn2009a.PhysRevLett.103.045301,
Kwon2014a.PhysRevA.90.063627,
Johnstone2019a,
Glidden:2020qmu,
Mueller2021a.PhysRevX.11.011053,
Mueller2024.PhysRevLett.132.094002,
Rasch2025bArxiv} 
and various types of wave turbulence
\cite{Zakharov1992a,
Nazarenko2011a,
Kozik2004a,
Mueller2006a,
Lvov2010a,
Boue2011a.PhysRevB.84.064516,
Navon2016a,
Navon2018a.Science.366.382},
both in theory and experiment.
While turbulence is often considered in the form of stationary cascades exhibiting invariance under spatial rescaling, it is inherently dynamic and, more generally, includes the possibility of scaling in time as well.
Such scaling characteristics far from equilibrium have been studied in analogy to equilibrium critical phenomena, as a manifestation of non-thermal fixed points  
\cite{Berges2008a,
Berges2008b,
Scheppach2009a,
Nowak2011a,
Schole2012a,
PineiroOrioli2015a,
Chantesana2018a,
Mikheev2018aNoArxiv,
Boguslavski:2019ecc},
see
\cite{Nowak:2013juc,
Berges2015a,
Schmied2018c,
Berges:2020fwq,
Siovitz2023b} 
for overviews.
This approach aims to eventually classify space-time scaling evolutions systematically based on fundamental principles such as symmetries and their manifestation in the prevalent configurations of quantum fields.
Recent experimental
\cite{%
Henn2009a.PhysRevLett.103.045301,
Gring2011a,
AduSmith2013a,
Kwon2014a.PhysRevA.90.063627,
Langen2015b.Science348.207,
Navon2015a,
Navon2016a,
Rauer2017a.arXiv170508231R.Science360.307,
Gauthier2019a,
Johnstone2019a,
Eigen2018a,
Prufer2018a,
Erne2018b,
Navon2018a.Science.366.382,
Glidden:2020qmu,
GarciaOrozco2021aNoArxiv,
Lannig2023a,
Martirosyan:2023mml,
Gazo2025a,
MorenoArmijos2024a,
Martirosyan2024a}
and theoretical studies 
\cite{%
Nowak2012b,
Berges2014a,
Berges2015b,
Moore:2015adu,
Nessi2014a.PhysRevLett.113.210402,
Gagel2014a.PhysRevLett.113.220401,
Bertini2015a.PhysRevLett.115.180601,
Babadi2015a.PhysRevX.5.041005,
Buchhold2015a.PhysRevA.94.013601,
Hofmann2014a,
Maraga2015a.PhysRevE.92.042151,
Williamson2016a.PhysRevLett.116.025301,
Williamson2016a.PhysRevA.94.023608,
Villois2016a.PhysRevE.93.061103,
Berges2016a, 
Bourges2016a.arXiv161108922B.PhysRevA.95.023616,
Cidrim2016a.PhysRevA.93.033651,
Chiocchetta:2016waa.PhysRevB.94.174301,
Karl2017b,
Schachner2016a,
Berges2017a,
Walz:2017ffj.PhysRevD.97.116011,
Deng2018a,
Baggaley2018a.PhysRevA.97.033601,
Bland2018a.PhysRevLett.121.174501,
Schmied2018d,
Mazeliauskas2018a,
Schlichting:2019abc,
Berges2019a,
Williamson2019a,
Gao2020a.PhysRevLett.124.040403,
Schmied:2018osf.PhysRevA.99.033611,
Schmied2019a,
Spitz2021a.SciPostPhys11.3.060,
Groszek2020a.SciPostPhys.8.3.039,
Groszek2021a.PhysRevResearch.3.013212,
Wheeler2021a,
Gresista2021a,
RodriguezNieva2021a,
Preis2023a.PhysRevLett.130.031602,
Liu2022a,
Mikheev:2022fdl, 
Heinen2023a,
Siovitz:2023ius.PhysRevLett.131.183402,
Noel2024:PhysRevD.109.056011,
Siovitz2025a,
Rosenhaus2024a.PhysRevE.109.064127,
Rosenhaus2024a.PhysRevLett.133.244002,
Rosenhaus2024b,
Rosenhaus2025a,
Noel2025a}, 
have explored various phenomena and properties of universal space-time scaling, largely in the field of ultracold atoms.
This includes (wave) turbulence, as well as coarsening and phase-ordering kinetics \cite{Bray1994a.AdvPhys.43.357,Puri2019a.KineticsOfPT,Cugliandolo2014a}, often in the context of multi-component systems.

The space-time scaling solutions near non-thermal fixed points are usually formulated for correlation functions in a given field theory.
These solutions describe self-similar transport of, e.g., the single-particle excitation population described by the momentum-space density $n(k,t)\sim t^{\alpha}f_\text{s}(t^{\,\beta}k)$, with a universal scaling function $f_\text{s}$, which 
typically shows a Porod-type power-law form $f_\text{s}(k)\sim k^{-\zeta}$.
If the associated transport is directed towards lower momenta, i.e., for $\beta>0$, it reflects a coarsening process. 
The exponents $\zeta$, $\beta$, and $\alpha$ are related to each other through conservation laws and have first been predicted analytically for the transport of Goldstone-type excitations in O$(N)$- and U$(N)$-symmetric scalar theories in the large-$N$ limit \cite{Berges2008a,
Berges2008b,
Scheppach2009a,
PineiroOrioli2015a,
Chantesana2018a,
Mikheev2018aNoArxiv,
Boguslavski:2019ecc,
Rosenhaus2024a.PhysRevLett.133.244002,
Rosenhaus2025a}.
In contrast, for low $N$, universal dynamics near non-thermal fixed points is usually dominated by excitations of the total density.
These include, in particular, nonlinear excitations such as vortices, solitons, or kinks 
\cite{%
Erne2018b,
Johnstone2019a,
Glidden:2020qmu,
GarciaOrozco2021aNoArxiv,
Lannig2023a,
Martirosyan:2023mml,
Gazo2025a,
MorenoArmijos2024a,
Martirosyan2024a,
Nowak2012b,
Schole2012a,
Moore:2015adu,
Williamson2016a.PhysRevLett.116.025301,
Williamson2016a.PhysRevA.94.023608,
Villois2016a.PhysRevE.93.061103,
Karl2017b,
Deng2018a,
Baggaley2018a.PhysRevA.97.033601,
Bland2018a.PhysRevLett.121.174501,
Schmied:2018osf.PhysRevA.99.033611,
Schmied2019a,
Groszek2020a.SciPostPhys.8.3.039,
Groszek2021a.PhysRevResearch.3.013212,
Wheeler2021a,
Spitz2021a.SciPostPhys11.3.060,
Liu2022a,
Siovitz:2023ius.PhysRevLett.131.183402,
Noel2024:PhysRevD.109.056011,
Siovitz2025a,
Noel2025a,
Rasch2025bArxiv}.
In that case, coarsening close to a non-thermal fixed point is reflected in the loss process of these excitations and thus in the growth of a characteristic length scale $\ell(t)\sim t^{\,\beta}$, associated with the mean inter-defect distance \cite{Schole2012a,Karl2017b}.

Here, we extend the analysis of non-thermal fixed points in a single-component Bose condensate in two spatial dimensions to a system with long-range dipolar interactions \cite{Lahaye2009a.ReptProgrPhys.72.126401, Chomaz2022a}, in addition to the local contact interaction.
In the superfluid regime, the partially attractive dipolar interaction leads to magnetostriction of trapped condensates \cite{ODell2004a,Eberlein2005.PhysRevA.71.033618,Wenzel2018a} and an anisotropic dispersion relation \cite{Santos2000a,Bismut2012a}, which induces anisotropic critical superfluid velocities \cite{Yu2017a,Ticknor2011a,Wenzel2018a}.
Under stronger confinement, the dispersion exhibits a non-monotonic dependence on momentum and thus a local, so-called roton minimum \cite{ODell2003a,Santos2003a,Chomaz2018a,Petter2019a}, which persists in quasi-2d \cite{Ticknor2011a} and eventually softens causing mean-field instabilities.
The dipolar interaction also intimately changes the properties of vortices \cite{Martin2017a}, which have been experimentally detected in \cite{Klaus2022a}.
In a quasi-2d geometry, the vortex core exhibits an elliptic shape for in-plane polarization 
and density ripples under the presence of a rotonic excitation \cite{Yi2006a,Ticknor2011a,Mulkerin2013a,Mulkerin2014a} which alter the dynamics of the vortices \cite{Mulkerin2014a,Gautam2014a}.

We consider a quasi-2d, dipolar Bose gas in the superfluid regime, away from the transition to a supersolid.
By preparing the system to exhibit coarsening of vortex ensembles and the buildup and decay of superfluid turbulence, we observe a universal space-time scaling evolution.
This enables us to assess the effect of the long-range and anisotropic interaction on the scaling dynamics near non-thermal fixed points.
For this we make use of the Gross-Pitaevskii equation including the contact and dipolar interactions between polarized dipoles of arbitrary polarization direction.
For both the idealized case of an \sludilute condensate, with large density and weak coupling, as well as for experimentally realistic systems, we find anomalously slow coarsening with a strongly sub-diffusive exponent, i.e., $\beta\approx1/5\ll1/2$, as in the non-dipolar case 
\cite{Karl2017b,
Deng2018a,
Johnstone2019a,
Spitz2021a.SciPostPhys11.3.060,
Noel2024:PhysRevD.109.056011,
Noel2025a,
Rasch2025bArxiv}.
Following this evolution, the transition to diffusion-type ($\beta\approx1/2$) coarsening is observed.
Unlike 3d simulations of dipolar gases \cite{Bland2018a.PhysRevLett.121.174501}, we find the above exponents to be universally valid across different scenarios.
In contrast to this universality across non-dipolar and dipolar gases, the vortices in the non-dipolar case tend to cluster and form large-scale eddies akin to turbulent flow, while in the dipolar systems, in particular with dipoles tilted into the 2d plane, the defects tend to anti-cluster and thus maximise distances between vortices of either sign. 

Our paper is organized as follows:
In Sect.~\ref{sec:quasi2d_dipolar_bose_gas}, we introduce the quasi-2d Gross-Pitaevskii model with dipolar interactions, which we solve within the Truncated-Wigner approach, using the parameters and initial far-from-equilibrium configurations discussed in Sect.~\ref{sec:initialization}, with a focus on the absence of mean-field instabilities.
In Sect.~\ref{sec:self_similar_scaling}, we observe the anomalous non-thermal fixed point in the self-similar scaling evolution of the single-particle occupation number spectra, with the scaling exponents being consistent with those of vortex coarsening as shown in Sect.~\ref{sec:coarsening_dynamics}.
The phenomenology of (anti)clustering of defects and its relation to sub-diffusive scaling are discussed in Sect.~\ref{sec:clustering}.
We summarize our findings and draw conclusions in Sect.~\ref{sec:conclusions}.

\section{Quasi-2d dipolar Bose gas}
\label{sec:quasi2d_dipolar_bose_gas}

Our goal is to study the universal dynamics of a single-component, dipolar Bose gas, in a quasi-2d geometry,  in the superfluid phase. 
We will evaluate its time evolution beyond mean-field, within the semi-classical Truncated-Wigner approach \cite{Blakie2008a}.
This means propagating initial configurations including randomly sampled quantum noise with the non-linear Gross-Pitaevskii equation (GPE) \cite{Baranov2002a,Baranov2008a, Blakie2009a,Ticknor2011a,Baranov2012condensed,Chomaz2022a},
\begin{align}
	\label{eq:gpe_3d}
	\i \hbar \pdv{\Psi(\vb{r},t)}{t} &= \left( - \frac{\hbar^2}{2m} \nabla^2 + V_\mathrm{ext}(\vb{r})+ g \rho(\vb{r},t) + \Phi_\mathrm{dd}(\vb{r},t) \right) \Psi(\vb{r},t) \,,
\end{align}
in the mean-field stable regime. $V_\mathrm{ext}(\vb{r})$ is an external trapping potential, $\rho(\vb{r},t) \equiv \abs{\Psi(\vb{r},t)}^2$ denotes the local 3-dimensional density of particles, and $m$ is their mass.
In three dimensions, the microscopic short-range interaction is approximated, at the prevalent low scattering energies, by a local contact interaction, with coupling constant $g=4\pi\hbar^2 a_\text{s}/m$ defined in terms of the $s$-wave scattering length $a_\text{s}$.
The long-range and anisotropic dipolar interactions,
\begin{align}
	\label{eq:dipolar_interaction_3d}
	\Phi_\mathrm{dd} (\vb{r}, t) &= \int \dd[3]{r'} \rho(\vb{r'}, t)  U_\mathrm{dd}(\vb{r}'-\vb{r}) \,,
\end{align}
are given in terms of the dipolar potential,
\begin{align}
	\label{eq:dipolar_potential_3d}
	U_\mathrm{dd}(\vb{r}) &= \frac{\Cdd}{4\pi} \frac{1-3\cos[2](\varphi)}{\abs*{\vb{r}}^3} \,,
\end{align}
with dipolar coupling $\Cdd$, depending on the distance vector $\vb{r}$ between the dipoles.
The interaction potential \eqref{eq:dipolar_potential_3d} is valid in the presence of a strong external magnetic field polarizing the magnetic atomic dipoles along a given axis $\hat{\mathbf{P}}$, and $\varphi$ denotes the angle between $\vb{r}$ and $\hat{\mathbf{P}}$.

The convolution theorem implies that the dipolar interaction is local in Fourier space, 
\begin{align}
	\label{eq:dipolar_interaction_3d_Fourier}
	\Phi_\mathrm{dd} (\vb{k}) &= \rho(\vb{k}) U_\mathrm{dd}(\vb{k}) \,,
\end{align}
with interaction potential
\begin{align}
	\label{eq:dipolar_potential_3d_Fourier}
	U_\mathrm{dd}(\vb{k}) = \Cdd \frac{3\cos[2](\varphi_{\vb{k}})-1}{3} \,.
\end{align}
Here, $\varphi_{\vb{k}}$ is the angle between $\vb{k}$ and $\hat{\mathbf{P}}$.

The dipolar Bose gas features mean-field instabilities for strong dipolar interactions, which are caused by the attractive force along $\hat{\mathbf{P}}$.
For the case of a vanishing external trapping potential $V_\mathrm{ext}=0$ in $d=3$, the chemical potential \(\mu = (1 - \epsdd) \, \rho g \) for a homogeneous density \(\rho(\vb{r}) \equiv \rho\) receives a correction to its non-dipolar value $\rho g$, which is proportional to the relative dipolar-to-contact interaction strength \(\epsdd = \Cdd/(3g)\).
For $\epsdd>1$ the long-wavelength modes with $k<\sqrt{2}k_{\xi}=\sqrt{2} / \xih$, with healing length $\xih = \hbar / \sqrt{2m \abs{\mu}}$, are unstable in mean-field approximation.
In this situation the gas may be stabilized by quantum fluctuation effects, which typically are, in leading order perturbation theory, described by the Lee-Huang-Yang (LHY) corrections \cite{Schuetzhold2006a,Lima2011a,Lima2012a,Waechtler2016a,Waechtler2016b,Bisset2016a}.
We do not include these corrections here and thus limit our analysis to the semi-classical TW simulations on the basis of the classical GPE.

Starting from the fully three-dimensional homogeneous system we implement an external trapping potential $V_\mathrm{ext}(\vb{r})= m \omega_z^2 z^2 /2$ that confines the system along the $z$-direction with trapping frequency $\omega_z$.
If the harmonic oscillator length
is smaller than the healing length, $\aho = \sqrt{\hbar/(m\omega_z)} \lesssim \xih$, the system is called quasi-2d, meaning that the dynamics along the $z$-direction is frozen out, i.e., to a good approximation in the single-particle ground state of the harmonic oscillator along the $z$-direction,
\begin{align}
	\label{eq:ground_state_harm_osc}
	h(z) = \left(\pi \aho^2\right)^{-1/4} \exp{-\frac{z^2}{2\aho^2}} \,.
\end{align}
This implies that the wave-function factorizes into planar and $z$-dependent parts, \(\Psi(\vb{r},t) = \psi(\vb{r}_\perp, t) h(z)\), and allows us to explicitly eliminate the $z$-dependence in \eqref{eq:gpe_3d}.
The resulting quasi-2d GPE reads
\begin{align}
	\label{eq:gpe_quasi2d}
	\i \hbar \pdv{\psi(\vb{r}_\perp, t)}{t} &= \left( - \frac{\hbar^2}{2m} \nabla_\perp^2 + g_{2} \rho_{2}(\vb{r}_\perp ,t) + \Phi_\mathrm{dd}^\perp(\vb{r}_\perp ,t) \right) \psi(\vb{r}_\perp,t) \,,
\end{align}
with local planar particle density $\rho_{2}(\vb{r}_\perp ,t) \equiv \abs{\psi(\vb{r}_\perp,t)}^2$ and renormalized contact interaction
\begin{align}
	\label{eq:coupling_renormalized}
	g_{2} = \frac{g}{\sqrt{2\pi} \aho} \,.
\end{align}
When integrating out the $z$-direction in \eqref{eq:dipolar_interaction_3d_Fourier} to obtain the interaction strength $\Phi_\mathrm{dd}^\perp(\vb{k}_\perp,t) = \rho_{2}(\vb{k}_\perp, t) U^\perp_\mathrm{dd}(\vb{k}_\perp)$ as a function of the planar momentum transfer $\vb{k}_\perp=(k_{x},k_{y})$, the harmonic trapping axis $\hat{\mathbf{e}}_{z}$ and the polarization axis $\hat{\mathbf{P}}$ of the magnetic dipole moments must be distinguished.
In Fourier space, the quasi-2d dipolar potential reads \cite{Fischer2006a,Ticknor2011a,Baillie2015a}: 
\begin{align}
	\label{eq:dipolar_potential_quasi2d}
	U_\mathrm{dd}^\perp(\vb{k}_\perp) 
	&= \frac{C_\mathrm{dd}}{3\sqrt{2\pi} \aho} 
	\qty[ F_\parallel (\tilde{k}_\perp, \tilde{k}_x) \sin[2](\theta) + F_\perp (\tilde{k}_\perp) \cos[2](\theta) ] \,,
\end{align}
where \(\tilde{k}_\perp \equiv k_\perp \aho / \sqrt{2}\), $\tilde{k}_x \equiv k_x \aho / \sqrt{2}$, and $k_\perp = \abs{\vb{k}_\perp}$.
The angle \(\theta=\measuredangle(\hat{\mathbf{e}}_{z},\hat{\mathbf{P}})\) is defined between the harmonic trapping along the \(z\)-axis and the polarization axis and will serve as a tuning parameter for the anisotropy of the interaction by tilting the dipole moments in the $x$-$z$-plane.

The auxiliary functions in \eqref{eq:dipolar_potential_quasi2d} are defined as
\begin{align}
	\label{eq:auxiliary_function_par}
	F_\parallel(\tilde{k}_\perp, \tilde{k}_x) &= 3 \sqrt{\pi} \frac{\tilde{k}_x^2}{\tilde{k}_\perp} \e^{\tilde{k}_\perp^2} \mathrm{erfc}(\tilde{k}_\perp) -1 \,,  \\
	\label{eq:auxiliary_function_perp}
	F_\perp(\tilde{k}_\perp) &= 2-3\sqrt{\pi} \tilde{k}_\perp \e^{\tilde{k}_\perp^2} \mathrm{erfc}(\tilde{k}_\perp) \,,
\end{align}
in terms of the complementary error function $\mathrm{erfc}(k)=1-\mathrm{erf}(k)$
\footnote{The asymptotic behavior $\mathrm{erfc}(k) \sim \mathrm{exp}(-k^2) / (\sqrt{\pi} k)$ for large values of $k$ can lead to arithmetic underflow when using numerical implementations of $\mathrm{erfc}$. By using instead implementations of the scaled complementary error function $\mathrm{erfcx}(k) = \mathrm{exp}(k^2) \mathrm{erfc}(k)$ one circumvents this. Further details can be found, e.g., at this \href{http://ab-initio.mit.edu/faddeeva/}{URL}}.

We introduce dimensionless coordinates 
\begin{align}
	\label{eq:coords_dimless}
	\tilde{x} = \frac{x}{\aho} \,, \quad \tilde{k} = k \aho \,, \quad \tilde{t} = \omega_z t \,,
\end{align}
as well as dimensionless fields and couplings
\begin{align}
	\label{eq:couplings_dimless}
	\tilde{\psi}(\tilde{\vb{r}}_\perp, \tilde{t}) = \psi(\vb{r}_\perp, t)\aho  \,, \quad 
	\tilde{a}_\text{s} = \frac{a_\text{s}}{\aho} \,, \quad 
	\tilde{a}_\mathrm{dd} = \frac{\add}{\aho} \,,
\end{align}
with the dipolar length denoted by $a_\mathrm{dd} \equiv \epsdd a_\text{s}$.
Inserting \eqref{eq:coords_dimless} and \eqref{eq:couplings_dimless} into \eqref{eq:gpe_quasi2d} yields the quasi-2d, dimensionless GPE
\begin{align}
	\label{eq:gpe_quasi2d_dimless}
	\i \pdv{\tilde{\psi}(\tilde{\vb{r}}_\perp, \tilde{t})}{\tilde{t}} &= \qty( - \frac{1}{2} \tilde{\nabla}_\perp^2 
	+ \sqrt{8\pi} \tilde{a}_\text{s} \tilde{\rho}_{2}(\tilde{\vb{r}}_\perp, \tilde{t}) 
	+ \tilde{\Phi}_\mathrm{dd}^\perp (\tilde{\vb{r}}_\perp, \tilde{t}) )\, 
	\tilde{\psi}(\tilde{\vb{r}}_\perp, \tilde{t}) \,.
\end{align}
The dimensionless chemical potential $\tilde\mu_{2}=\mu_{2}/(\hbar \omega_z)$, in two dimensions, in mean-field approximation results as
\begin{align}
	\label{eq:chemical_potential_mf}
	\tilde{\mu}_{2} (\theta)= \sqrt{8 \pi} \tilde{a}_\text{s} \tilde{\rho}_{2} \left[1+\epsdd \left(3 \cos^2 \theta - 1 \right) \right] \,,
\end{align}
in terms of which the healing length in the quasi-2d system reads
\begin{align}
	\label{eq:healinglength2d}
	\xih 
	&= \frac{\hbar}{\sqrt{2m \abs{\mu_{2}}}}
	= \frac{a_\text{ho}}{\sqrt{2\abs{\tilde{\mu}_{2}}}}
         \,.
\end{align}
We are going to drop the tildes and the index $\perp$ in what follows.

\section{Simulations of far-from-equilibrium turbulent dynamics}
\label{sec:initialization}
In this section we present our simulation scheme of far-from-equilibrium universal dynamics of a quasi-2d dipolar gas in the superfluid phase.
Starting from a range of initial field configurations containing ensembles of quantum vortices, the system is found to enter a regime, which exhibits a superfluid turbulent flow and its coarsening evolution marked by an algebraic decay of the vortex density in time due to vortex-antivortex annihilation processes.

\subsection{Two different parameter sets}
\label{subsec:parameters}
We define two different parameter sets, which we denote as \sludilute and \slquantum cases. 
In the \udilute case, we follow Ref.~\cite{Karl2017b} and consider a system with a very small diluteness parameter such that the rotonic nature of the dipolar dispersion as well as higher-order fluctuations are irrelevant.
In the \slquantum case, we choose experimentally realistic gas parameters with smaller interparticle distance relative to the scattering length.
Here the anisotropy induced by the dipolar interaction matters and can lead to a rotonic spectrum. 
Our simulations operate in the regime of high mode occupancies, i.e., in a semi-classical regime, where we can apply the truncated Wigner (TW) approximation.
Since we do not include the stabilizing LHY correction in the GPE \eqref{eq:gpe_quasi2d_dimless}, we restrict ourselves to the mean-field stable, superfluid region of the phase diagram \cite{Mulkerin2014a,Hertkorn2021a.PhysRevResearch.3.033125}.

\subsubsection{Ultradilute gas}
\label{subsec:parameters_ud}

For the non-dipolar case,  we start by choosing the same values for the particle number, $N = \rho_2  L^2 = 3.2 \cdot 10^9$, and the dimensionless 2d coupling strength, $\tilde{g}_{2}=g_{2}m/\hbar^{2} = (\mu_2 / \rho_2)(m/\hbar^{2}) = 1.5 \cdot 10^{-5}$, as in \cite{Karl2017b}, where, moreover, the mass was set to $m=1/2$. 
Choosing a linear system size $\tilde{L}=L/\aho \approx 1612$ in units of the harmonic oscillator length in $z$-direction \footnote{In our calculations, we originally used a different set of units, which lead to the particular value for $L$.}, we ensure that $ \xi_\mathrm{h}/\aho=(2\abs{\tilde{g}_{2}N})^{-1/2}\tilde{L}\approx5$, and thus that the superfluid is effectively two-dimensional.

At the transverse peak density, this choice corresponds to an extraordinarily small 3d diluteness parameter 
\begin{align}
  \eta=\sqrt{\rho_{3} a_\text{s}^{3}}
  =\frac{\sqrt{N\tilde{g}_{2}^{3}}}{2^{1/4}4\pi\,\tilde{L}}
  =\frac{\tilde{g}_{2}\aho}{2^{3/4}4\pi\,\xi_\text{h}}
  \simeq1.4\cdot10^{-7}\,.
\end{align}
Hence, the system is ultradilute, and, thus, increasing $\eta$of the 2d superfluid requires increasing $\tilde{g}_{2}$ while keeping $\xi_\text{h}/\aho$ sufficiently larger than 1.

The above parameters result in a mean planar density $\rho_{2} = N / L^2 \approx 1232 \, \aho^{-2}$, and a dimensionless chemical potential
\begin{align}
	\label{eq:chemical_potential_value}
  	\tilde\mu_{2} = {\tilde{g}_2\rho_2\aho^{2}} \approx 0.0184 \,.
\end{align}
In order to obtain comparable results for different dipolar-to-contact interaction strengths $\epsdd$, and for a clearer distinction of the effects of tilting the polarization to $\theta>0$, we fix the chemical potential for any $\epsdd$ and $\theta=0$ to the value in \eq{chemical_potential_value}, $\tilde\mu_{2}(\epsdd, \theta=0) = \tilde\mu_{2}$.
We explore the anisotropy of the dipolar interaction potential \eqref{eq:dipolar_potential_quasi2d} by repeating our simulations for nonzero tilting angles, $\theta \in \{\pi/8, \pi/4\}$, which thus modifies the chemical potential.

As a consequence of fixing $\mu_{2}$ for $\theta=0$, the scattering length $\tilde{a}_\text{s}$, for a given dipolar strength $\epsdd$, is given by
\begin{align}
	\label{eq:scattering_length_from_mu}
	\tilde{a}_\text{s} &= \frac{\tilde{\mu}_{2}}{\sqrt{8\pi} \tilde{\rho}_{2} (1+2\epsdd)} \,.
\end{align}
For $\epsdd=0$, this gives $a_\text{s}\simeq3\cdot10^{-6}\,\aho$ and, for a realistic harmonic oscillator length, $\aho\simeq0.25\,\mathrm{\mu m}$ (for $^{164}$Dy and $\omega_z =2\pi \cdot 1\,\mathrm{kHz}$), a 3d peak density $\rho_{3}\approx\, 4.4 \cdot 10^4 \mathrm{\mu m}^{-3}$, and a scattering length $a_\text{s}\simeq7.5 \cdot 10^{-4}\, \mathrm{nm}$.
Hence, the \udilute gas has a higher 3d density and much smaller scattering length as compared with experimentally realistic values for ultracold atomic gases,
$a_\text{s} \approx5 \, \mathrm{nm}$, $\rho_3=100\dots1000\,\mathrm{\mu m}^{-3}$, cf.~\Subsect{parameters_q}.
Moreover, due to the small value of $\tilde\mu_{2}$, the ensuing dynamics is dominated by the kinetic term in \eqref{eq:gpe_quasi2d_dimless}, and the interactions take little effect.

\subsubsection{Dysprosium gas: quantum regime}
\label{subsec:parameters_q}
For the \quantum case, we choose parameters reflecting experimentally realistic systems, cf. \cite{Chomaz2022a}, specifically, a length $L=48.4 \, \aho$, density $\rho_{2}=42.8 \, \aho^{-2}$, and chemical potential $\mu_{2} = 16.2 \, \hbar \omega_z$.
This results in a dipolar length equal to that of \ce{^164 Dy} in the case of $\epsdd = 1.47$, cf.~\cite{Maier2015.PhysRevA.92.060702,Tang2015.PhysRevA.92.022703}.
As tilting angles we choose again $\theta \in \{0,\pi/8,\pi/4\}$.
For $\epsdd=0$, the above values give $\tilde{g}_{2}=\tilde{\mu}_{2}/\tilde{\rho}_{2}\approx0.379$, $a_\text{s}\simeq0.08\,\aho$, a diluteness parameter $\eta=(\rho a_\text{s}^{3})^{1/2}\simeq0.1$, and, for a typical harmonic oscillator length, $\aho\simeq0.25\,\mathrm{\mu m}$, to a 3d density of $\rho_{3}\approx1.5\cdot10^3\,\mathrm{\mu m}^{-3}$.

For all parameter settings, the healing length, $\xih = \hbar / \sqrt{2m \abs{\mu}}=\aho/\sqrt{2\tilde{\mu}_{2}}$, can be resolved by the numerical lattice spacing. 
In the \quantum case, it results as $\xih\approx0.17\,\aho$, hence, violating the quasi-2d condition $\xih\gg\aho$.
We can therefore not expect to quantitatively describe the full three-dimensional dynamics using a two-dimensional simulation.
However, in order to feature rotonic excitations on the order of the healing length which lead to density ripples around vortices, cf.~\Fig{clustering_single_run}, and eventually instabilities, cf.~\Fig{instab_diag}, we require the system to be in a crossover regime from quasi-2d to 3d \cite{Fischer2006a,Lahaye2009a.ReptProgrPhys.72.126401}.
In this regime of strongly oblate traps, the dominant vortex dynamics remains in-plane and Kelvin-wave excitations are strongly suppressed \cite{Rooney2011a}, hence, vortex dynamics is still sufficiently described using the quasi-2d approximation.
We leave the quantitative comparison between a full 3d and a quasi-2d simulation for future work.

\subsection{Dipolar instabilities and anisotropy of interactions}
\label{subsec:mf_instabilities}

The GPE \eqref{eq:gpe_quasi2d_dimless} exhibits Bogoliubov quasiparticle excitations with dispersion
\begin{align}
	\label{eq:Bog_dispersion}
	\omega(\kvperpq)^2 &= \epsilon_{\kperpq} \left[ \epsilon_{\kperpq} + 2 \rho_2 \left( g_\mathrm{2} + U_\mathrm{dd}^\perp(\kvperpq) \right) \right] \,,
\end{align}
with single-particle energy $\epsilon_{\kperpq} = \kperpq^2 /2$ which turns anisotropic for tilting angles $\theta>0$ \cite{Fischer2006a,Ticknor2011a}.
Increasing the dipolar strength and/or tilting the polarization into the $x$-$y$-plane changes \eqref{eq:Bog_dispersion} from the monotonic, non-dipolar dispersion towards a non-monotonic dispersion exhibiting a local maximum (maxon excitation) and a local minimum (roton excitation), cf.~\Fig{instab_diag}.
This reduces the stabilizing effect of the tight harmonic confinement along the $z$-axis and ultimately gives rise to instabilities, which, in mean-field approximation, are associated with $\omega(\kvperpq)$ turning imaginary for certain modes.
One distinguishes phonon instabilities, where -- for a particular orientation of \(\kvperpq\) -- all momenta below a threshold \(\kperpq<k_\mathrm{ph}\) turn unstable, and roton instabilities, where this is the case for modes in a range \(k_{\mathrm{r},1}<\kperpq<k_{\mathrm{r},2}\) of non-vanishing wave numbers \cite{Santos2003a,Fischer2006a, Baranov2008a,WilsonRM2008a,WilsonRM2009a,Blakie2012a,Mulkerin2014a,Chomaz2022a}.
Since we simulate \eqref{eq:gpe_quasi2d_dimless} without LHY corrections, we need to avoid these mean-field instabilities in order to obtain physically meaningful results.
For a given dipolar strength $\epsdd$, this sets a limit to the possible tilting angles.
All parameters defined in the previous section are chosen such as to avoid instabilities, also those caused by beyond-mean-field fluctuations appearing during the dynamics far from equilibrium, as discussed in more detail in \App{mf_instabilities}.

We find that the vortex patterns and clustering of same-sign defects come out very differently in the non-dipolar and dipolar cases, as well as for different tilting angles, which we expect to relate to the different character of excitations created in the system by the vortex-antivortex annihilation processes.
An interesting question thus concerns the values of the respective healing-length and roton scales and the isotropy of the elementary excitation energies in momentum space.
We discuss these questions in more detail in \App{krotheal}.
This shows that roton and healing momenta are of the same order in the \quantum case, lying in a region of momenta where also the dispersion of elementary excitations shows its strongest deviations from that in a non-dipolar gas.
In contrast, in the ultradilute case, the roton momentum is more than an order of magnitude smaller than the healing momentum and the modification of the dispersion due to the dipolar interactions is less significant.
We furthermore find that the anisotropy caused by the tilting affects mainly the excitations close to the healing momentum, below which one typically observes universal coarsening dynamics after a strong quench.

\subsection{Initial states and numerical procedure}
\label{subsec:inistates}
%

\begin{figure*}[t]
	\centering
	\includegraphics[width=\textwidth]{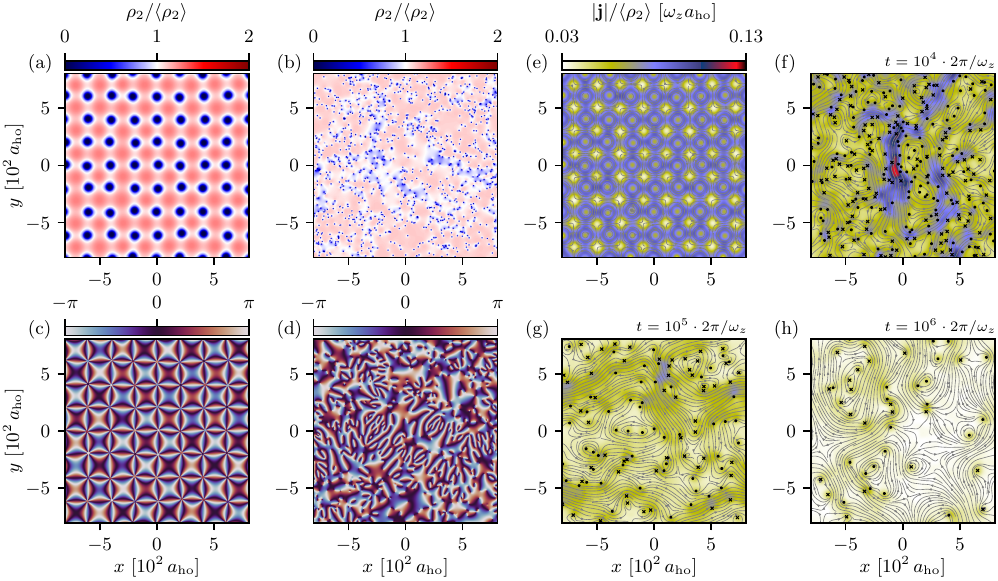}
	\caption{
		(a,b) Initial density $\rho_2/ \langle \rho_2 \rangle$ and (c,d) phase $\varphi=\arg \psi$ for both (a,c) lattice and (b,d) random sampling after a phase imprint and a brief evolution in imaginary time for the vortex cores to form.
		Panels (e--h) show the normalized, absolute current $|\vb{j}| = |\rho_2 \nabla \varphi|$ in the \udilute case for $\epsdd=0.5$ and $\theta=0$, starting from the lattice initial condition.
		The gray streamlines indicate the local strength and direction of the current field $\vb{j}$ and the dots and crosses for $t>0$ show the positions of the vortices and antivortices, respectively.
		The temporal evolution is shown for the initial configuration at (e) $t=0$, as well as at (f) $t=10^4 (2\pi)/\omega_z$, (g) $t=10^5 (2\pi)/\omega_z$, and (h) $t=10^6 (2\pi)/\omega_z$.
		In particular at early times we observe high flows encircling clusters of vortices which coarsen over time leading to a weaker flow and smaller vortex clusters.
	}
	\label{fig:initial_cond}
\end{figure*}

In order to assess the dependence of the dynamics on the chosen initial condition, we adopt different far-from-equilibrium initial field configurations both containing a large number of vortices.

In the first setting, `lattice sampling', of the \udilute system, we imprint phase defects with winding number $q=\pm6$ in an $8\times8$ lattice into a uniform background density.
This is done in a checkerboard manner, i.e., alternating between vortices and antivortices.
To accelerate decay of these vortices to elementary defects with $q=\pm1$, a random small offset of the vortices' positions from the checkerboard grid is added.
As shown in \cite{Karl2017b}, from this initial vortex configuration, the system can approach a non-thermal fixed point corresponding to anomalously slow coarsening, i.e., mutual annihilation and thus spatial dilution of the defects.
For comparison, we further employ `random sampling', where we imprint $1000$ elementary ($q=\pm1$) phase defects with vanishing total angular momentum at random positions into a uniform background.
In the \quantum regime, for $\epsdd=1.47$, we prepare of $4\times4$ lattices with winding $q=\pm4$ and reduce the number of randomly sampled vortices to $100$ in order to avoid mean-field instabilities.

After imprinting the phase patterns, a brief evolution in imaginary time allows the vortex cores to form, thus avoiding shock waves in the subsequent early-time evolution.
Since we do not explicitly pin the vortices for lattice sampling, the higher-order winding number vortices already decay into a very small cluster of $|q|$ elementary defects throughout the imaginary time propagation, cf.~\Fig{initial_cond}(c).
The two, lattice and random, initial configurations are shown in \Fig{initial_cond}  (a--d), in density and phase.

The systems' evolution is then computed on a $1024^2$ numerical lattice using a split-step Fourier algorithm in real time, up to a maximum time $t_\mathrm{max} = 10^6  (2\pi)/\omega_z$ with time step  $\dd{t} = 0.1  (2\pi)/\omega_z$ (\udilute case) and $t_\mathrm{max} = 10^3  (2\pi)/\omega_z$ with $\dd{t} = 10^{-4}  (2\pi)/\omega_z$ (\quantum case).
By the choice of a pseudo-spectral algorithm we introduce periodic boundary conditions to our system, which must already be obeyed by the initial configuration.

Besides single evolution runs, we evaluate time evolving correlation functions within the semi-classical Truncated-Wigner approximation \cite{Blakie2008a}, by adding initial noise in the form of half a quasiparticle occupation with random noise in the homogeneous-density Bogoliubov modes and averaging our results over either $50$ (ultradilute) or $100$ (quantum) runs from different such initial configurations.
For numerical efficiency the simulations are performed in a parallelized fashion using GPU clusters.

\subsection{Turbulent flow and vortex pattern coarsening}
\label{subsec:vortexdyn}

In panels (e--h) of \Fig{initial_cond}, starting from lattice initial conditions, an exemplary time evolution is shown, in the \udilute case for $\epsdd=0.5$ and $\theta=0$.
The four snapshots encode the momentary magnitude of the current field $\vb{j}= \rho_2 \vb{v} = \rho_2 \nabla \theta$ in the 2d volume, together with gray streamlines and arrows indicating the direction of the fluid flow.
The initial lattice of $q=\pm6$ vortices -- which enter in the form of very small vortex clusters -- quickly breaks up into elementary defects, and strong coherent flow prevails at early times, with only few annihilations taking place, cf.~panel (f) at $t=10^4 (2\pi)/\omega_z$.
After the initial reordering phase, the regions of strong flow encircle clusters of vortices which break up throughout the dynamics into smaller clusters surrounded by a decelerated flow, cf.~(g) $t=10^5(2\pi)/\omega_z$ and (h) $t=10^6(2\pi)/\omega_z$.
Hence, starting with the vortex ensembles described in the previous section, the system quickly exhibits a turbulent flow pattern, which is characterized by vortex scattering and mutual vortex-antivortex annihilation, and thus by coarsening of the superfluid flow and vortex pattern, which gradually weakens the flow and leads back to an equilibrium state.

In case of random initial conditions the initial reordering phase appears absent and the system almost immediately enters the universal scaling dynamics.
Videos of the different evolutions studied are provided online \cite{VideosDipolarAnomNTFP}.

In the next section our focus will be, at first, on the universal dynamical properties of the intermediate stage of the equilibration process.

\section{Space-time scaling of momentum spectra}
\label{sec:self_similar_scaling}
Universal space-time scaling, in field theory, is, in principle, observed in correlation functions, which are averaged over many different runs.
We begin by analyzing the self-similar scaling evolution of single-particle occupation-number spectra during the vortex pattern coarsening in the superfluid, as described above.
We compare various dipolar with a non-dipolar system, for the parameters defined in \Subsect{parameters}.

\subsection{Scaling form of the momentum spectrum}
\label{subsec:observables}

The angle-averaged single-particle spectrum
\begin{align}
	\label{eq:single_particle_spectrum}
	n(k,t) &= \int  \dd{\Omega} \expval{\psi^*(\kvperpq, t) \, \psi(\kvperpq, t)} \,,
\end{align}
with $k=\abs{\kvperpq}$ and $\text{d}\Omega$ denoting the vector's surface angle measure represents such a correlation function.
As mentioned in the introduction, close to a non-thermal fixed point the dynamics of correlation functions is described by a universal scaling form,
\begin{align}
	\label{eq:occupation_number_scaling}
	n(k,t) &= (t/t_0)^\alpha n([t/t_0]^\beta k, t_0) 
	\equiv(t/t_0)^\alpha f_\text{s}(k/k_{\Lambda}(t))\,,
\end{align}
with universal scaling function $f_\text{s}(k/k_{\Lambda}(t_{0})) = n(k,t_0)$ and reference time $t_0$.
The exponents, $\alpha$ and $\beta$, determine the scaling dimensions of $f_\text{s}$ and of the IR momentum scale $k_{\Lambda}(t)=(t/t_0)^{-\beta} k_{\Lambda}(t_{0})$ with respect to time.
This also sets the scaling dimension of the characteristic length scale
\begin{align}
	\label{eq:avg_idd_scaling}
	\ell(t) \sim k_{\Lambda}(t)^{-1}\sim (t/t_0)^{\beta} \,,
\end{align}
which will be discussed further in Sect.~\ref{sec:coarsening_dynamics} in terms of the mean separation between defects and its algebraic growth in time, reflecting the coarsening process of the system.

In the following, we focus on the self-similar evolution in the infrared (IR), in which the evolution represents an inverse flow of particles towards lower momenta, resulting in $d$ spatial dimensions in the scaling relation \cite{PineiroOrioli2015a,Chantesana2018a},
\begin{align}
	\label{eq:exponents_particle_conservation}
	\alpha=d \, \beta\,.
\end{align}
Within this IR region, in general after a possible `prescaling' stage \cite{Schmied2018d}, the scaling function is found to be well approximated by the scaling form \cite{Karl2017b}
\begin{align}
	\label{eq:scalingfunction}
	f_\text{s}(\kappa)
	&= \frac{A}{1+\kappa^{\zeta}}\,,
\end{align}
such that $n(k,t)$, for $k\ll  k_{\Lambda}(t)$ shows a plateau while it falls off with a Porod-like tail $n(k,t)\sim k^{-\zeta}$ for $k\gg  k_{\Lambda}(t)$.

We found that our imprints of random vortex configurations into a condensate with uniform density, within a system with periodic boundary conditions, did not ensure that the total momentum of the system vanished for each realization.
While it is expected to vanish on average, it can be non-zero in the single realizations, remaining conserved in time due to spatial translation invariance of the Hamiltonian.
We therefore rather define the angle-averaged single-particle spectrum as 
\begin{align}
	\label{eq:single_particle_spectrum_shifted_mean}
	n(\kvperpq,t)&= \int \dd{\Omega} \expval{\psi^*(\kvperpq - \expval{\kvperpq}, t) \, \psi(\kvperpq - \expval{\kvperpq}, t)} \,,
\end{align}
i.e., as the average of the Galilei-shifted single realizations, such that each such realization has a vanishing net total momentum.
We found that this yields a significantly better matching of the resulting average momentum distribution by the scaling form defined in \eq{occupation_number_scaling} and \eq{scalingfunction}, in particular for the case of initial random vortex sampling and strong dipolar interactions and polarization tilting.

\subsection{Example results of scaling dynamics}
\label{subsec:scaling_collapse}
%

\begin{figure*}[t]
	\centering
	\includegraphics[width=\textwidth]{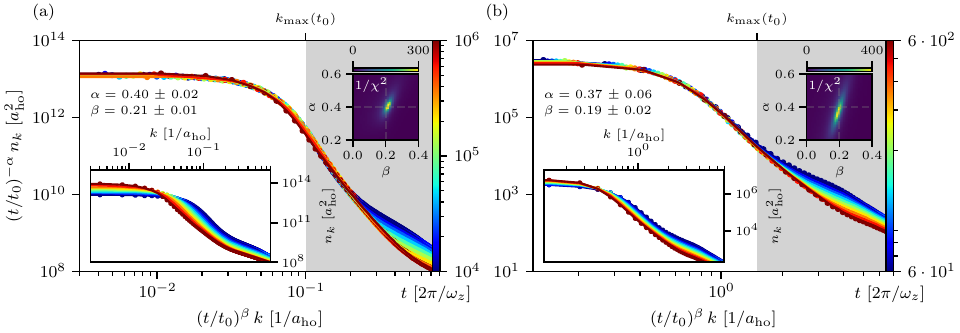}
	\caption{
		Rescaled occupation number $n(k,t)$ in the (a) \udilute regime for $\epsdd=0.5$ and (b) \quantum regime for $\epsdd=1.47$, both for $\theta=0$, lattice sampling and over two, respectively one, order of magnitude in time.
		The gray, ultraviolet (UV) region is excluded from the rescaling procedure in order to identify the IR scaling behavior, such that the exponents are extracted by fitting the scaling collapsed distributions within the momentum window $k\in[0,k_\text{max}(t)]$ only, see main text for details.
		We obtain $\alpha=0.40\pm0.02$ and $\beta=0.21\pm0.01$ in the \udilute and $\alpha=0.37\pm0.06$ and $\beta=0.19\pm0.02$ in the \quantum case.
		The non-rescaled density spectra are shown in the lower-left inset for comparison.
		In (a) at $t=10^5  (2\pi)/\omega_z$ we extract the spatial exponent $\zeta=5.44\pm0.09$ and in (b) at $t=10^2  (2\pi)/\omega_z$ the exponent $\zeta=4.39\pm0.10$, cf.~\eq{occupation_number_scaling}, \eqref{eq:scalingfunction}.
		The inverse of the deviation $\chi^2$ of the least-squares fit for the respective scaling exponents is shown in the upper-right inset and indicates the optimal scaling exponents.
	}
	\label{fig:rescale_occ_number}
\end{figure*}

\Fig{rescale_occ_number} (a) shows the self-similar scaling evolution of $n(k,t)$, for a system with dipolar interaction strength $\epsdd=0.5$, polarization into the $z$-direction, $\theta=0$, and initial lattice sampling while all other parameters have been chosen as defined in Sect.~\ref{subsec:parameters_ud} for the \udilute regime.
The inset in the lower-left displays a set of snapshots of the momentum spectra within the time interval $t\in[10^{4}\dots10^{6}](2\pi/\omega_{z})$, which have been rescaled, for the main panel, to demonstrate the self-similarity between different times.
The snapshots are logarithmically spaced in order to equally weight all times in the rescaling procedure.
The collapse onto a single scaling function shown is achieved with the exponents 
$\alpha=0.40\pm0.02$ and $\beta=0.21\pm0.01$.
By fitting the scaling function \eqref{eq:scalingfunction} at $t=10^5  (2\pi)/\omega_z$ we extract $\zeta=5.44\pm0.09$.

For the rescaling procedure according to \eqref{eq:occupation_number_scaling}, only values in the momentum range $k\in[0,k_\text{max}(t)]$, with $k_\text{max}(t)=(t/t_{0})^{-\beta}k_\text{max}(t_{0})$, $t_{0}=10^{4}  (2\pi)/\omega_{z}$, $k_\text{max}(t_{0})=0.1\,a_\text{ho}^{-1}$ are included (white, non-shaded area in the main panels of \Fig{rescale_occ_number}).
The distribution of the inverse of the deviations $\chi^2$ from the least-squares fit 
in the $\alpha$-$\beta$ plane exhibits the optimal overlap of occupation numbers and thus the exponents (upper-right insets).
We adopted here the rescaling procedure outlined in the appendix of \cite{PineiroOrioli2015a}.
The extracted exponents are consistent within errors with $\beta=0.193\pm0.05$, $\alpha=0.402\pm0.05\approx d\beta$, and $\zeta=5.7\pm0.3$, as reported in \cite{Karl2017b} for sub-diffusive anomalously slow scaling of a single-component, \udilute non-dipolar Bose gas with contact interactions in $d=2$ dimensions.

Qualitatively, we find the same scaling in the \quantum regime, see \Fig{rescale_occ_number} (b) for the dysprosium parameter set, with $\epsdd=1.47$, $\theta=0$, lattice sampling and all other parameters as defined in \Subsect{parameters_q}.
We will discuss the details and differences between the \udilute and \quantum regime in the subsequent sections.

Before proceeding to this we remark on the form of the momentum spectrum $n_{k}$ obtained in the UV, which corresponds to a Rayleigh-Jeans distribution $\sim T/\epsilon_{k}$.
We can quantify, from this, the kinetic energy per particle at the latest time, once most defects have decayed and energy has been redistributed into the UV.
In the ultradilute regime in \Fig{initial_cond} (h) we obtain $E_\mathrm{kin}/N \approx 0.004 \,\hbar \omega_z$ and for lattice sampling in the quantum regime with \(\epsdd=1.47\) and \(\theta=0\) we get $E_\mathrm{kin}/N \approx 0.6\, \hbar \omega_z$.
In both cases the imprinted energy remains only a fraction of the system's ground state energy given by the respective chemical potential, cf.~Sect.~\ref{subsec:parameters}.
To estimate the temperature of the nearly equilibrated UV tail, we compare the Rayleigh-Jeans law with the occupation number spectrum \eqref{eq:single_particle_spectrum_shifted_mean} in the UV above the healing scale, obtaining $k_\mathrm{B} T \sim 10\,\hbar\omega_z$ (ultradilute) and $k_\mathrm{B} T \sim 1 \,\hbar \omega_z$ (quantum).
For the TWA to remain accurate, the largest single-particle energy $\epsilon_{k_\mathrm{max}}$ ought to remain $\lesssim k_\mathrm{B} T$ and occupancies not to fall below $\sim1/2$, which holds in the ultradilute regime ($\epsilon_{k_\mathrm{max}}\approx4\,\hbar\omega_z$) but is violated in the quantum regime ($\epsilon_{k_\mathrm{max}}\approx4000\,\hbar\omega_z$).
Hence, in the latter case we do not expect to correctly describe the thermalisation behavior as well as the correlations in the UV; however, for the redistribution of particles in the IR and the vortex motion the repercussions are expected to be minor.
We have explicitly verified that energy continously flows into the UV modes and enables the appearance of anomalously slow scaling also for a larger grid constant, which forces particles to remain at lower energies.
The level of ultraviolet excitations may affect, though, the point of time when the scaling starts and ends to appear, see the discussion in the subsequent sections.
Computing the dynamics for the low particle densities under consideration in the quantum case, by taking into account quantum fluctuations, could be achieved by means of extended techniques such as PGPE or stochastic GPE simulations, see, e.g., \cite{Blakie2009a,Rooney2012a.PhysRevA.86.053634,Rooney2014a,Cockburn2013qgft.conf,Proukakis2025a}, which we leave for future work.

\subsection{Time dependence of exponents}
\label{subsec:time_dependence_exponents}
%

\begin{figure*}[t]
	\centering
	\includegraphics[width = \textwidth]{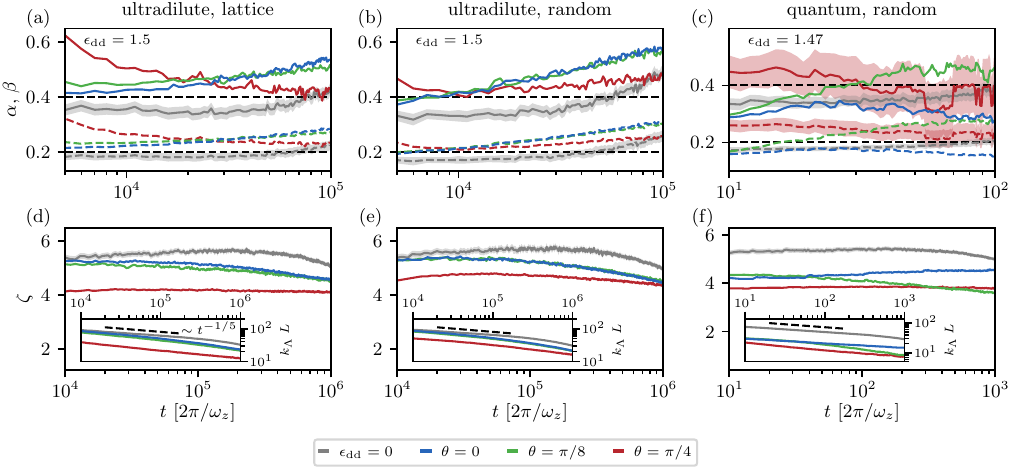}
	\caption{
		Time-varying scaling exponents $\alpha(t)$ (solid) and $\beta(t)$ (dashed) are shown for different tilting angles in the \udilute case at $\epsdd=1.5$ for (a) lattice and (b) random sampling and (c) in the \quantum case for $\epsdd=1.47$ and random sampling.
		Panels (d--e) show the respective evolutions of the exponent $\zeta(t)$.
		The exponents $\alpha$ and $\beta$ are obtained by collapsing $10$ momentum spectra, logarithmically spaced in time within $[t,10t]$, onto each other.
		As in \Fig{rescale_occ_number}, the exponents are extracted by fitting the scaling collapsed distributions within the momentum window $k\in[0,k_\text{max}(t)]$, where the initial cutoff $k_\text{max}(t_0)$ is fixed for the smallest and largest initial time and interpolated to intermediate initial times $t_0$ using a power law. 
		The gray shadow displays the error obtained for the scaling exponents of the non-dipolar case; its magnitude is of similar order for the other settings in the \udilute regime (not shown).
		For the strongly tilted case in (c) the error is indicated as a red shadow, which demonstrates its increase in the \quantum regime.
		In (a) and (b) the extracted exponents are consistent with $\beta=1/5$ and $\alpha=d\beta$, with a trend to higher values at late times, which is more pronounced for randomly sampled initial vortex configurations, cf.~the main text.
		For (c) we also find consistent exponents within errors, we observe however stronger trends in the variations of the exponents.
		The exponent $\zeta$ is extracted by fitting the scaling function \eqref{eq:scalingfunction} to the momentum spectra. 
		The UV cutoff is fixed both at the earliest and latest times and interpolated by a power law for intermediate times.
		The extracted non-dipolar exponents are consistent with $\zeta=5.7\pm0.3$ found in \cite{Karl2017b} for most times.
		In general, a constant exponent $\zeta$ is found in all \udilute scenarios until $t\approx 3\cdot10^5$ before a tendency of decreasing exponents indicates the transition of the scaling function towards the diffusion-like scaling, cf.~the main text.
		In the \quantum regime $\zeta$ is constant for most times, besides for the weakly-tilted $\theta=\pi/8$ case.
		The insets show the characteristic momentum scale, which exhibits power-law scaling $k_\Lambda(t)\sim t^{-1/5}$ in the universal regime of constant $\zeta$.
	}
	\label{fig:running_rescale_occ_number}
\end{figure*}

The scaling exponents obtained in \Fig{rescale_occ_number} are only for one particular interval in time and for specific parameters of the dipolar strength and tilting angle.
In order to assess the stability and longevity of the observed scaling behavior, we show
in \Fig{running_rescale_occ_number} (a--c) the scaling exponents $\alpha(t)$ (solid) and $\beta(t)$ (dashed), as functions of time $t$, obtained from collapsing $10$ spectra onto each other -- chosen logarithmically spaced 
within $[t,10\,t]$ -- where $t$ is chosen in different intervals as seen in the different panels.
\Fig{running_rescale_occ_number} (d--f) shows the corresponding evolution of the exponent $\zeta(t)$ and in the insets the characteristic momentum scale $k_\Lambda(t)$, both obtained by fitting the scaling function \eqref{eq:scalingfunction} to the evolving momentum spectra.

\begin{figure*}[t]
	\centering
	\includegraphics[width = \textwidth]{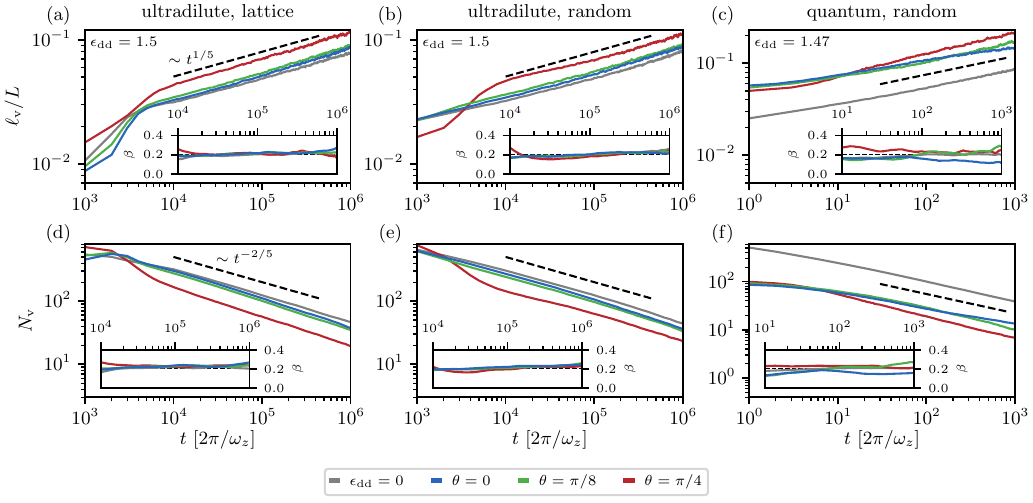}
	\caption{%
		In panels (a--c) the average inter-defect distance $\ell_\mathrm{v}$ and in (d--f) the vortex number $N_\mathrm{v}$ are shown for (a,d) lattice and (b,e) random sampling in the \udilute regime at $\epsdd=1.5$ and (c,f) in the \quantum regime for random sampling, all for different tilting angles.
		For comparison, the corresponding non-dipolar $\epsdd=0$ (gray) case is shown in all panels, respectively.
		The scaling in (a--c) with $\beta=1/5$ and in (d--f) with $-2\beta=-2/5$ of the anomalous non-thermal fixed point is indicated by the black-dashed lines and qualitatively follows the observed scaling at late times.
		In the insets the dynamically fitted scaling exponent $\beta$ is plotted where the shaded area in the insets indicates the error of the fitting procedure.
	}
	\label{fig:avg_idd}
\end{figure*}

The scaling exponents are shown in both the \udilute and \quantum regimes, for different dipolar strengths and tilting angles, as indicated in the panels.
For comparison also the non-dipolar (gray) results are displayed in each panel.

For both sampling schemes in the \udilute regime, panels (a) and (b), the extracted exponent $\beta(t)$ shows consistency with a constant anomalous value $\beta(t)\simeq\beta\approx1/5$ found in a non-dipolar system \cite{Karl2017b}, taking into account the extracted errors.
The same applies for the exponent $\alpha\approx 2/5$, corroborating the relation $\alpha=d\beta$ corresponding to total particle-number conservation in the self-similar transport.
Similar results have been found for $\epsdd=0.5$ (cf.~\App{scal_exp_05}) and $\epsdd=1$ (not shown).
Nevertheless, for all dipolar strengths and tilting angles, a trend to larger exponents is found at late times which is more pronounced for random sampling.
This has also been observed in the non-dipolar case \cite{Karl2017b}, where the system was found to transit to diffusion-type scaling with $\beta=\alpha/d\simeq0.5$ at late times, which was attributed to background noise dominating the mutual annihilation of vortices, cf.~also \Sect{coarsening_dynamics}.
The onset of this transit appears to happen at earlier times for increasing relative dipolar interaction strength and for small tilting angles, compare here $\epsdd=0$ and $1.5$.
This was also observed with $\epsdd=0.5$ (cf.~\App{scal_exp_05}) and $1$ (not shown).

For the \quantum regime in \Fig{running_rescale_occ_number} (c) we find scaling exponents of $\beta\approx0.2$ for the non-dipolar and the isotropic dipolar case over the whole time-range covered, showcasing the presence of the anomalous non-thermal fixed point also in the \quantum regime.
In the weakly tilted case $\theta=\pi/8$, the system also starts off at $\beta\approx0.2$ but departs at intermediate times towards larger scaling exponents, while in the strongly tilted case $\theta=\pi/4$, the system approaches anomalous scaling rather at later times, and with larger fluctuations.
These can be attributed to the plateau of the scaling function \eqref{eq:scalingfunction} being deeply in the IR at late times and thus limiting the rescaling procedure according to \eqref{eq:occupation_number_scaling} which consequently leads to increased errors on $\alpha$ as is shown by the red shadow.
For smaller tilting angles this effect turns out to be less pronounced and therefore results in smaller errorbars.
The relation $\alpha\approx d\beta$ is found to be well satisfied except in the strongly tilted case.

For lattice sampled initial configurations in the \quantum regime we obtained similar results (not shown).
We further computed, as discussed in \App{anisotropy_spectrum}, scaling exponents \eqref{eq:occupation_number_axis_averaged_scaling} based on the separate rescaling of the average occupation numbers \eqref{eq:single_particle_spectrum_axis_averaged} along the cartesian spatial directions.
They did not exhibit any qualitative difference compared to the radial scaling exponents and obeyed the relation $\alpha_{x,y} \approx \beta_{x,y}$.

For early times $t<3\cdot10^5$ a relatively constant $\zeta$ in \Fig{running_rescale_occ_number} indicates universal scaling, which is corroborated by the power-law scaling $k_\Lambda(t)\sim t^{-1/5}$ of the characteristic momentum shown in the insets of (d--f).
For (d) lattice and (e) random sampling at $\epsdd=1.5$ in the \udilute regime, $\zeta(t)$ changes at late times $t\gtrsim3\cdot10^5$, as the system departs towards the diffusion-type scaling (cf.~Sect.~\ref{subsec:coarsening_late_time}).
The late-time change of $\alpha$ and $\beta$, as found in (a) and (b), can be attributed to the deterioration of the scaling function \eqref{eq:scalingfunction}.
Within the universal regime we find agreement of the non-dipolar $\zeta$ with the result obtained in \cite{Karl2017b}.
However, for increasing tilting angle we observe a significant reduction towards the prediction $\zeta=d+2=4$ \cite{Nowak2011a} for a random defect distribution while still observing anomalously slow coarsening both in $k_\Lambda(t)$ and $\beta(t)$.
This is also present in the \quantum regime (f) and for weaker dipolar strengths (not shown), however, less pronounced.
These results suggest that the exponent $\zeta$ is less universal than the scaling exponent $\beta$ and primarily determined by the defect configurations on scales in between $1/\ell \lesssim k \lesssim 1/\xi_\mathrm{h}$, which relates to clustering discussed in \Sect{clustering}.
Also Kelvin-wave excitations \cite{Noel2025a} could modify this exponent.

In the \quantum regime in \Fig{running_rescale_occ_number} (f) we find $\zeta(t)$ to be relatively stable for the dipolar curves with $\theta=0$ and $\theta=\pi/4$, in agreement with their corresponding stable $\beta(t)$ shown in panel (c).
Both for the non-dipolar and the weakly tilted dipolar cases a decreasing $\zeta(t)$ at late times marks the departure from the anomalous scaling as can also be seen in the steeper scaling of $k_\Lambda(t)$ at late times.

\section{Vortex pattern coarsening}
\label{sec:coarsening_dynamics}
A central question of pattern coarsening dynamics concerns the relation between the particular way the pattern evolves and decays in time and the shape and scaling dynamics of the momentum spectrum $n(k,t)$ \cite{Mikheev2023a.EPJST232.3393}.
Specifically, as $\beta$ determines the algebraic decay of the characteristic wave number scale $k_{\Lambda}$ and thus the growth of the length scale $\ell(t)\sim k_{\Lambda}(t)^{-1}\sim t^{\,\beta}$, it is expected to also characterize the coarsening of the vortex pattern, such that the average inter-defect distance grows as  $\ell_\mathrm{v}(t)\sim t^{\,\beta}$.
Using a plaquette algorithm which computes the phase winding of the phase field $\varphi=\arg \psi$ around all $2\times2$ plaquettes on the grid, we identify the positions of the vortices at each moment in time from the non-zero windings around the respective plaquettes, which proves sufficient for the purpose of the scaling analysis.

From this, we determine the mean intervortex distance $\ell_\mathrm{v}$ by averaging over the distances between each pair of nearest-neighbor vortices, also across the periodic boundaries.
As the increase of the average inter-defect distance is due to mutual annihilation of vortices and antivortices, also the total number of vortices is determined as a function of time.

In \Subsect{coarsening_anomalously_slow} we analyse the initial stage of anomalously slow coarsening, where $\beta\approx1/5$, which is followed by a transition to diffusive scaling with $\beta\approx1/2$, discussed in \Subsect{coarsening_late_time}. 
In \Subsect{driven_dissipative} we show how the transition time depends on the friction present in the system.
In \Sect{clustering} we study the (anti)clustering behavior throughout the universal regime.

\subsection{Anomalously slow scaling of the intervortex distance}
\label{subsec:coarsening_anomalously_slow}

\Fig{avg_idd} shows the evolution of the characteristic scale $\ell_\mathrm{v}$ and the vortex number $N_\mathrm{v}$ for different configurations, together with the corresponding evolution of the coarsening exponent $\beta(t)$, obtained by fitting power laws to $\ell_\text{v}(t)$ and $N_\text{v}(t)$ over half an order of magnitude in time, symmetrically around $t$, at logarithmically spaced points in time.

In the \udilute regime (\Fig{avg_idd} (a,b,d,e)), we observe that, irrespective of the initial condition and parameters, the characteristic scale and the vortex number exhibit anomalously slow power-law coarsening, with $\beta\approx1/5$ and $-2\beta\approx-2/5$, respectively, over almost two orders of magnitude up to the latest times.
At early times, $t\lesssim10^4\,(2\pi)/\omega_{z}$, a significantly smaller inter-defect distance is found for lattice sampling (cf.~\Fig{avg_idd} (a,d)), which quickly rises during early evolution times.
This is due to the clusters of vortices with $q=\pm1$ emerging from the break-up of the higher-order winding number ($q=\pm6$) vortices the initial vortex lattice consists of.
Similar results have also been found for weaker dipolar strengths at $\epsdd=0.5$ and $\epsdd=1$ (not shown).

At later times, the absolute value of the inter-defect distance and the vortex number are found to be in quantitative agreement between the isotropic case ($\theta=0$) and the non-dipolar case independent of the dipolar strength and the initial vortex configuration.
This independence of $\epsdd$ is due to our choice of a fixed chemical potential in the isotropic case, cf.~\Subsect{parameters_ud}, which implies equal healing lengths and thus similar characteristics of the vortices as well as linear sound excitations.
For the case of strongly-tilted ($\theta=\pi/4$) dipoles, quantitative deviations from the isotropic case are expected due to the different chemical potential \eq{chemical_potential_mf} and thus healing-length scale, and due to the anisotropic terms in \eq{auxiliary_function_par}.
The latter is however subdominant in the \udilute regime, cf.~\Fig{instab_diag}.

In the \quantum case, Figs.~\fig{avg_idd} (c,f) show scaling exponents $\beta\approx0.2$ independent of $\theta$ similar to the \udilute case.
In general, the evolution close to the anomalous non-thermal fixed point appears to be less stable than in the \udilute case which can be seen from the shorter universal regime and its stronger dependence on the tilting angle.
This can be expected to be caused by the lower fluid density in the \quantum regime, such that the effect of compressible excitations on the annihilation of nearby vortices is enhanced.
The quantitatively different inter-defect distances in the dipolar and the non-dipolar cases are due to a different number of initially sampled vortices in order to avoid mean-field instabilities, cf.~\Subsect{parameters_q}.

Hence, even more consistently than from the analysis of the spectra in \Sect{self_similar_scaling}, we find $\ell_\text{v}(t)\sim t^{\,\beta}$ and $N_\text{v}\sim t^{-2\beta}$ to a good approximation, with a constant sub-diffusive scaling exponent $\beta\approx0.2$.

To refine the analysis further, we also evaluated the time evolution of the average distances separately for the two spatial dimensions, cf.~\App{anisotropy_spectrum}. 
Within errors, we find isotropic scaling in the $x$-$y$-plane, as can be expected from the fact that the dipolar interactions \eq{dipolar_potential_quasi2d}, at leading order in $\kperpq$, as well as the ensuing dispersion in the sound-wave region, $\kperpq\ll k_\text{rot}$, scale the same way in both, $x$ and $y$, directions.
Anisotropies appear in the non-leading momentum dependence only, at least away from the transition to the mean-field unstable and thus supersolid phase.

\subsection{Transition to diffusive scaling of the intervortex distance}
\label{subsec:coarsening_late_time}
%

\begin{figure*}[t]
	\centering
	\includegraphics{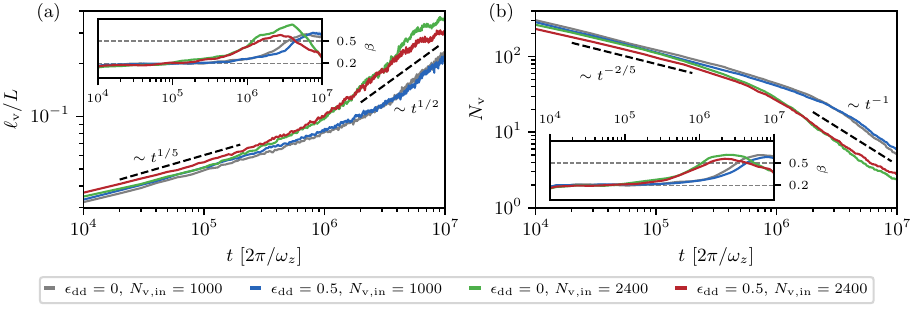}
	\caption{%
		(a) The average inter-defect distance for the non-dipolar ($\epsdd=0$) and the isotropic dipolar ($\epsdd=0.5$) case in the \udilute regime starting from random vortex sampling of either $1000$ or $2400$ vortices.
		The evolution time is extended by an order of magnitude in time until almost no vortices are left in the system.
		Starting form $2400$ vortices, we observe the transition towards the diffusion-type scaling with $\beta\approx1/2$ at $t\approx10^6  (2\pi)/\omega_z$ which is shifted to $t\approx4\cdot10^6  (2\pi)/\omega_z$ for an initial vortex number $N_\mathrm{v,in}=1000$.
		The two black-dashed lines serve as a guide to the eye for power-law scaling with $\beta=1/5$ and $\beta=1/2$.
		(b) The corresponding average number of vortices is shown for the same scenarios and exhibits the same transition from sub-diffusive towards diffusive coarsening behavior.
		The black-dashed lines indicate the power-law scalings with exponents $-2/5$ and $-1$.
		The insets show the dynamically fitted scaling exponent $\beta$ extracted from (a) the average inter-defect distance and (b) the average vortex number.
		They exhibit a transition between scaling exponents of $\beta\approx0.2$ towards $\beta\approx0.5$ at late times.
	}
	\label{fig:late_time_coarsening}
\end{figure*}

As the scaling exponents tend to drift away at late times, we investigate longer evolution times.
In \Fig{late_time_coarsening} (a), the long-time evolution of the average inter-defect distance is shown for the non-dipolar ($\epsdd=0$) and the isotropic dipolar ($\epsdd=0.5$, $\theta=0$) cases in the \udilute regime, while \Fig{late_time_coarsening} (b) shows the corresponding total number of defects.
In both evolutions we start with random vortex samples, cf.~\Fig{avg_idd}, starting with $1000$ or $2400$ initial vortices.
The latter increases the initial density of the defects involved to induce more background fluctuations during the anomalous scaling interval and gives rise to an earlier crossover time.
For $t\lesssim10^6$ we find scaling as in \Fig{avg_idd} (b), with scaling exponent $\beta\approx0.2$, while at late times $t\gtrsim10^6$, a transition to a diffusion-type scaling exponent $\beta\approx0.5$ is observed, and the transition time depends non-universally on the number of vortices in the initial state.
We, in fact, find the average vortex number to be smaller when starting with $2400$ vortices, cf.~\Fig{late_time_coarsening} (b), with even a few single runs, both dipolar and non-dipolar, having zero defects at the latest time.
This implies that the initial vortex distribution undergoes a strong vortex annihilation phase at early times, introducing background excitations that shift the transition to a diffusive behavior to earlier times.
As an aside, the annihilation of all vortices was also achieved starting from $1000$ defects in a single run at very late times $t>10^7 2\pi/\omega_z$.
The respective dynamically fitted, time-dependent exponents are shown in the insets.

As a side note, the observed transition between two distinctly different scalings could be associated with the picture developed for renormalization-group flow close to equilibrium fixed points: 
Generically, such flows cannot be fine-tuned to lie on a pure attractor leading into a fixed point but most likely evolve at most close to it, such that, after a finite period of critically slowed scaling, the flow departs again the vicinity of the fixed point and may then go over to approach a different fixed point in the space of system configurations.

\subsection{Coarsening subject to friction}
\label{subsec:driven_dissipative}

In the following, we explore more systematically the reasons for the transition from sub-diffusive to diffusion-type scaling.
A vortex-antivortex pair travels, in a pure, zero-temperature condensate, to a good approximation as a Helmholtz dipole, i.e., with propagation direction perpendicular to the dipole vector and without changing the intervortex separation.
Hence, pairs travel in parallel and their mutual annihilation is thus, in principle, impossible.
This constraint can also be traced back to the absence of a kinetic term in the Onsager Hamiltonian \cite{Onsager1949b} of point vortices, which takes the form of a Coulomb potential in two dimensions. 
For more than two vortices, only scattering between different such dipoles and vortices can lead to a modification of the mutual intervortex distances.

If, however, friction between the vortices and the background condensate is present due to the interaction with sound-wave and single-particle excitations, also the interdefect distance in a pair can shrink.
A dissipative force on the vortex adds a Magnus-type force perpendicular to the propagation direction, inducing vortex pairs of opposite circulation to approach and eventually annihilate, while equal-sign pairs further separate from each other.
For idealized, point vortices, i.e., for a vanishing healing length, these effects are captured by different terms in their Hall-Vinen-Iordanskii (HVI) equations of motion \cite{Hall1956a.PRSLA.238.204,Iordansky1964a.AnnPhys.29.335,Iordanskii1966JETP...22..160I}, which define the dissipative generalization of the Onsager dynamics.
As a result of this, the loss rate is proportional to the average number of defects present, $\partial_{t}{N_\mathrm{v}}\sim-N_\mathrm{v}^{2}$, such that $N_{\mathrm{v}}\sim t^{-1}$ decreases as a power-law in time, corresponding to an increase of the mean separation of defects as $\ell_\mathrm{v}(t)\sim t^{\,\beta}$ with $\beta=1/2$.

While this dissipation-induced decay is dominated by two-body processes, sub-diffusive scaling with $\beta\ll1/2$ requires a loss rate which scales higher than linear in $N_\mathrm{v}$.
If friction is subdominant, as pointed out above, vortices can approach each other solely through collisions of a weakly bound pair and a vortex, which allow the outgoing pair to be much more closely bound than the incoming pair.
Once a sufficiently closely bound pair emerges from such a collision, it can undergo mutual annihilation due to the remaining friction with the background, before it scatters and thus potentially increases its binding length again.
As a result, loss and thus coarsening is rather dominated by such collisions between pairs and vortices, and it can be argued that a rate corresponding to three-vortex collisions gives $\beta=1/5$ \cite{Karl2017b}.

\begin{figure}[t]
	\centering
	\includegraphics[width=\columnwidth]{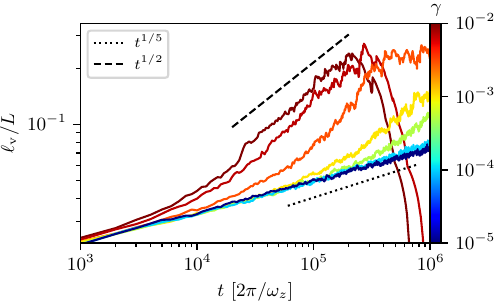}
	\caption{%
		Average inter-defect distance $\ell_\text{v}(t)$ for $\epsdd=0.5$, $\theta=0$, and random sampling of $1000$ defects in the \udilute case for different damping parameters $\gamma \in [10^{-5}, 10^{-2}]$ of the driven-dissipative GPE at $T=10.4 \, \hbar \omega_z / k_\mathrm{B}$.
		With increasing $\gamma$ the late-time diffusive scaling with $\beta\approx0.5$ sets in earlier. 
		When the final pair left in the system annihilates, $\ell_\text{v}$ drops to zero.
		The anomalous (dotted) and diffusion-like (dashed) power-laws are shown as a guide to the eye.
		The simulations have been performed on a $1024^2$ numerical grid which obeys the TWA condition $k_\mathrm{B} T \gtrsim \epsilon_{k_\mathrm{max}} \approx 4\hbar\omega_z$.
	}
	\label{fig:driven_diss_avg_idd}
\end{figure}

In the following we add noise to the dipolar gas in a controlled manner, by means of a Keldysh-type damping term proportional to the single-particle Hamiltonian \cite{Cockburn2012a},
\begin{align}
	\label{eq:driv_diss_gpe}
	\i \partial_t \psi &= (1-\i \gamma) \qty( - \frac{1}{2} \nabla^2 + \sqrt{8\pi} a_\text{s} \abs{\psi}^2 + \Phi_\mathrm{dd}^\perp ) \, \psi + \eta \,.
\end{align}
The dimensionless forcing term, which is subject to white-noise correlations, 
\begin{align}
	\label{eq:forcingcorr}
	\expval{\eta^*(\vb{x},t) \eta(\vb{x}',t')} 
	= 2\gamma \bar{T} \delta(\vb{x}-\vb{x}')\delta(t-t')\,,
\end{align}
determines the dimensionless temperature $\bar{T} =  k_\mathrm{B} T /(\hbar \omega_z)$ of the corresponding thermal bath.

In \Fig{driven_diss_avg_idd} the inter-defect distance is shown for an evolution according to \Eq{driv_diss_gpe} at a fixed temperature $T=10.4\,\hbar \omega_z / k_\mathrm{B}$, for different $\gamma \in [10^{-5},10^{-2}]$.
In comparison with the $\sim t^{1/5}$ (black-dotted) scaling at early times, the inter-defect distance approaches diffusion-type scaling with $\beta\approx0.5$ (black-dashed) the earlier the higher $\gamma$.
This observation matches qualitatively the behavior observed in non-dipolar gases \cite{Karl2017b}.
For the largest values of $\gamma$ we observe an eventual decrease of the inter-defect distance at late times when, on average, only two vortices are left, being mutually attracted due to the dissipative forces in the system.

\begin{figure*}[t]
	\centering
	\includegraphics[width=\textwidth]{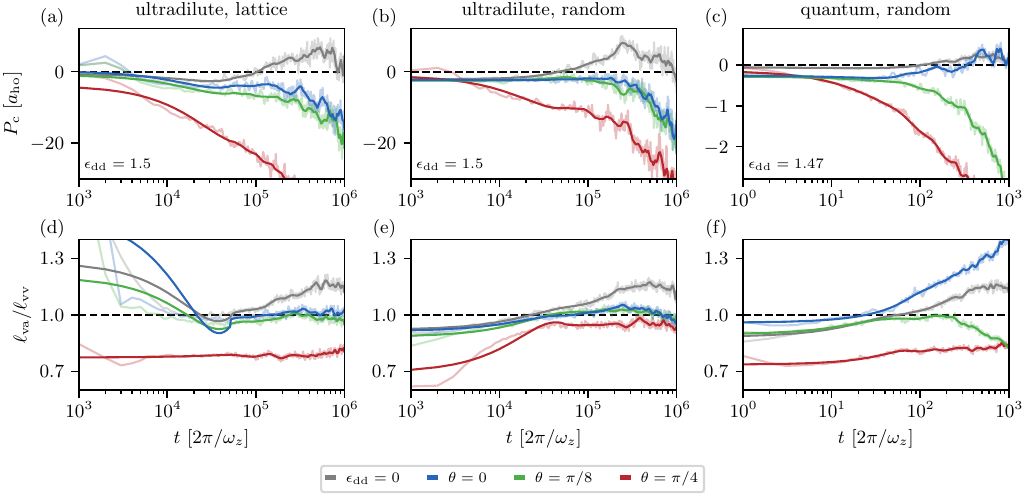}
	\caption{
		In panels (a--c) the degree of clustering $P_\mathrm{c}(t)$ and in (d--f) the ratio of the mean nearest-neighbor opposite-sign $\ell_\mathrm{va}$ and  equal-sign $\ell_\mathrm{vv}$ inter-defect distance are shown for (a,d) lattice and (b,e) random sampling in the \udilute case, with $\epsdd=1.5$ and (c,f) in the \quantum regime for random sampling, all for different tilting angles.
		The transparent lines show the fully time-resolved observables, while the solid ones are obtained by smoothing with a third-order Savitzky-Golay filter in order to highlight the tendency of (anti)clustering.
		$P_\mathrm{c}>0$ and $\ell_\mathrm{va} / \ell_\mathrm{vv}>1$ indicate clustering, whereas $P_\mathrm{c}<0$ implies anticlustering, i.e., maximized distances between equal-sign defects, and $\ell_\mathrm{va} / \ell_\mathrm{vv}<1$ hints towards the prevalence of vortex dipoles.
		}
	\label{fig:clustering}
\end{figure*}

\subsection{Vortex clustering}
\label{sec:clustering}

Based on direct analysis and the conjectured role of three-vortex collisions
it was proposed that clustering of like-sign vortices, which could suppress mutual annihilation of vortex dipoles by shedding vortices from opposite-sign partners could be a relevant precondition for anomalous scaling with $\beta\approx1/5$ \cite{Karl2017b}.
In the following, we analyse clustering and investigate modifications due to dipolar long-range and anisotropic interactions.

\subsubsection{Degree of clustering $P_\mathrm{c}$}
\label{subsec:degree_of_clustering}

Following \cite{Karl2017b} we measure the degree of clustering $P_\mathrm{c}$ of a given distribution of \(N_{\text{v}}^+ (t)\) vortices and $N_{\text{v}}^- (t) = N_{\text{v}}^+ (t)$ antivortices at time $t$, within a total volume $V$, through Ripley's function,
\begin{align}
	\label{eq:ripleys_function}
	\mathcal{L}(\ell,t) &= \qty( \frac{V}{\pi N_{\text{v}}^+(t)^2} \sum_{i\neq j} \Theta(\ell-\ell_{ij}(t)) )^{1/2} \,,
\end{align}
which counts for every vortex $i$ with positive winding number the number of same-sign vortices $j$ with relative distance $\ell_{ij}(t)\leq\ell$, taking into account the periodic boundary conditions.
Assuming a random defect distribution, every vortex has on average $\pi \ell^2 N_\mathrm{v}^+(t)/V$ neighboring same-sign defects within a circle of radius $\ell$.
Performing the summation over all positive-sign vortices accounts for another factor of $N_\mathrm{v}^+(t)$, thus resulting in $\mathcal{L}(\ell,t) = \ell$.
Hence, deviations of $\mathcal{L}(\ell,t) - \ell$ from zero signal (anti)clustering at a given scale.
While clustering leads, on average, to an overcount of vortices within a given area around any vortex, anticlustering, i.e., maximization of distances counts less than the mean.
These deviations imply
a degree of clustering of
\begin{align}
	\label{eq:deg_of_clustering}
	P_\mathrm{c}(t) &= \int_a^b \dd{\ell} \frac{\mathcal{L}(\ell,t)-\ell}{b-a} \,,
\end{align}
within a chosen range of scales between $a$ and $b$.
While $P_\mathrm{c}(t)=0$ indicates random sampling, a clustered (anticlustered) distribution implies $P_\mathrm{c}(t)>(<) \ 0$.
The lower integration bound $a$ is set to $\sim 2 \,\xih$ corresponding to the minimal distance between two vortices.
For the upper integration bound, half the system size, $b=L/2$, is chosen in order to avoid overlap of the circle around a vortex with itself across the periodic boundaries.

In \Fig{clustering} (a--c), the time-evolution of $P_\mathrm{c}(t)$ is shown for the same parameter sets as in \Fig{avg_idd}.
Comparing with our results for the anomalous scaling exponent, we find only for the non-dipolar case  in the \udilute regime a correlation of anomalous scaling and clustering at late times, corroborating \cite{Karl2017b}.
In contrast, for the dipolar systems, in particular with dipole tilting, anticlustering is found and increases with $\epsdd$ (not shown).
Similar results are seen in the quantum case, with a tendency to clustering found for the non-tilted case.
Hence, the correlation between subdiffusive scaling and clustering of same-sign vortices seen in the non-dipolar case is not preserved under dipolar interactions. 

Vortex clustering in single runs is discussed in more detail in \App{clustering_single_runs}.
Considering the spatial pattern of the 2d density $\rho_{2}$ around and in between the vortices, one sees that in the dipolar -- in particular for non-zero tiltings -- and in the quantum case, strong density ripples can be seen at later times, which are aligned with the tilting ($x$-)direction.
These ripples can be at least partially associated with the roton excitations, which are concentrated around the roton momentum scale, cf.~\Fig{instab_diag}, panels (b) and (d) \cite{Mulkerin2014a}.
For the quantum case with $\epsdd=1.47$ and strong tilting, the entire system, at early times, is filled with the weak stripe-shaped density ripples, cf.~\Fig{clustering_single_run} (f).
This suggests that the vortex-antivortex annihilation process in the dipolar system is different from that in the non-dipolar case and that many roton-like excitations are created, which influence the further evolution of the system.
Our results suggest, in particular, that these excitations prevent the vortex flow to develop large-scale eddies which give rise to clustering of same-sign vortices.
This is further underlined by the ratios between inter-defect distances of vortices of either opposite or same sign of circulation as evaluated in the following.

\subsubsection{Opposite- vs. equal-sign pair size}
\label{subsec:clustering_inter_defect_distance}

The function $P_\mathrm{c}(t)$ \eqref{eq:deg_of_clustering}, defined in the previous section, only differentiates between different degrees of (anti)clustering within a same-sign vortex configuration.
In order to refine our analysis we distinguish opposite-sign, $\ell_\mathrm{va}$, and equal-sign, $\ell_\mathrm{vv}$, inter-defect distances, which are defined accordingly. 
Their ratio $\ell_\mathrm{va} / \ell_\mathrm{vv}$ is shown in \Fig{clustering} (d--f) for the parameter sets used in \Fig{clustering} (a--c).

A ratio of $\ell_\mathrm{va} / \ell_\mathrm{vv}\approx1$, as found for the isotropic and weakly-tilted dipolar cases in the \udilute regime, indicates a random distribution. 
In contrast $\ell_\mathrm{va} / \ell_\mathrm{vv}>1$, as found for the non-dipolar case, indicates the formation of clusters of (at least two) equal-sign vortices and their mutual separation.
From the ratio $<1$ for the strongly tilted case we can infer that the anticlustered vortices tend to form vortex-antivortex dipoles.
In the case of weaker dipolar strengths $\epsdd$ we recover the same qualitative but less pronounced behavior (not shown).
In the \quantum case at late times, the non-dipolar and isotropic dipolar system exhibit ratios $>1$, indicating clustering, whereas the tilted, dipolar cases have ratios $<1$, indicating vortex dipoles to be dominant.
Hence, we find overall agreement with the (anti)clustering behavior inferred from \Fig{clustering} (a--c) complemented with the tendency of vortex dipole formation.

\section{Conclusions}
\label{sec:conclusions}

We investigated anomalous non-thermal fixed points in a dipolar, quasi-2d Bose gas, extending previous studies on contact-interacting Bose gases.
The polarization direction of the dipoles allows us to modify the anisotropy of the interaction by tilting the dipoles towards the transverse plane. 
The relative strength of the long-range interactions is controlled by means of $\epsdd$.

We, in general, recover the sub-diffusive scaling exponent $\beta\approx0.2$ characterising the coarsening of the vortex ensemble and the rescaling of single-particle momentum spectra of the superfluid.
We conclude that the anomalous non-thermal fixed point exhibits universal properties independent of the microscopic details of the solely contact or also dipolar interaction.
We could also confirm that the anomalous non-thermal fixed-point remains a partial attractor in the dipolar systems. 
At later times, faster coarsening follows the initial slow one, with a diffusion-type exponent $\beta\approx0.5$.
The transition time can be varied by the choice of initial conditions or by adjusting the strength of damping in a driven-dissipative setting.
Hence, anomalous scaling survives the longer, the less dissipative interaction between vortices and sound excitations prevails, which gives rise to a Magnus-type attractive force between the defects that accelerates their mutual pairwise annihilation.

As a new result, which goes beyond the known dynamics in a contact-interacting superfluid, the anomalously slow coarsening in the dipolar systems appears to not rely on the clustering of vortices.
Our data rather points to the opposite, i.e. a weak anti-clustering of defects and a dominance of vortex-antivortex pairs.
Moreover, these results provide a clear case, where the temporal exponents $\alpha$ and $\beta$ and thus the dynamical scaling is not tied to a unique spatial scaling function $f_\text{s}(k\gg k_{\Lambda})\sim k^{-\zeta}$ and thus to a unique spatial exponent $\zeta$.

Our findings emphasize the need for analytic characterisation of the observed attractors to establish universality classes related to non-thermal fixed points and identify their defining characteristics.
Within the mean-field treatment employed in this work, increasing the dipolar strength or the tilting angle can ultimately cause instabilities, which could lead to the collapse of the Bose gas 
\cite{Dodd1996a,
Sackett1998a,
Roberts2001a.PhysRevLett.86.4211,
Goral2002a,
ODell2003a,
Santos2003a,
Giovanazzi2004a.EPJD.31.439,
Ronen2006a,
Ticknor2008a,
Koch2008stabilization,
Parker2009a,
Metz2009coherent}.
The condensation of strongly magnetic atoms in experiment 
\cite{Beaufils2008a,
Lu2011a,
Aikawa2012a} 
has demonstrated, however, that this mean-field instability is counteracted by corrections beyond mean-field order \cite{Schuetzhold2006a,Lima2011a,Lima2012a,Waechtler2016a,Waechtler2016b,Bisset2016a}, which stabilize the system and give rise to a supersolid phase where the condensate exhibits periodic density modulations \cite{Tanzi2019a,Chomaz2019a,Boettcher2019a} or takes the form of isolated droplets 
\cite{Kadau2016a,Chomaz2016,Schmitt2016self}.
Given that our work focused on the superfluid regime of a dipolar condensate, extending this analysis of far-from-equilibrium dynamics to the supersolid regime and assess the effect of broken translational invariance on the scaling exponents will be of great interest.
Irrespective of universal dynamics, we observed a strong dependence of the (anti)clustering behavior on the dipolar interaction which asks for a better understanding of the phenomenology of (Onsager)-clustering in dipolar systems.

\begin{acknowledgments}
The authors thank S.M. Roccuzzo for inspiring discussions and collaboration in the early stages of the present work.
They thank K. Chandrashekara, L. Falzoni, P. Heinen, W. Kirkby, H. K{\"o}per, A.-M. Oros, and I. Siovitz for discussions and collaboration on related topics.
The authors are grateful to D. Proment for point out the role of roton excitations in the vortex annihilation process.
The authors acknowledge support 
by the Deutsche Forschungsgemeinschaft (DFG, German Research Foundation), through 
SFB 1225 ISOQUANT (Project-ID 273811115), 
grant GA677/10-1, 
and under Germany's Excellence Strategy -- EXC 2181/1 -- 390900948 (the Heidelberg STRUCTURES Excellence Cluster), 
and by the state of Baden-W{\"u}rttemberg through bwHPC, the data storage service SDS{@}hd supported by the Ministry of Science, Research and the Arts Baden-W{\"u}rttemberg (MWK), and the DFG through grants 
INST 35/1503-1 FUGG, INST 35/1597-1 FUGG, and INST 40/575-1 FUGG
(SDS, Helix, and JUSTUS 2).
\end{acknowledgments}


\vspace{\columnsep}

\begin{appendix}
\begin{center}
\textbf{APPENDIX}
\end{center}
\setcounter{equation}{0}
\setcounter{table}{0}
\makeatletter

\section{Limitations to parameter choices}

In this appendix we discuss our parameter choices with respect to mean-field instabilities and effects of anisotropy due to tilting.

\subsection{Avoiding instabilities due to attractive interactions}
\label{app:mf_instabilities}

\begin{figure*}[t]
	\centering
	\includegraphics[width=\textwidth]{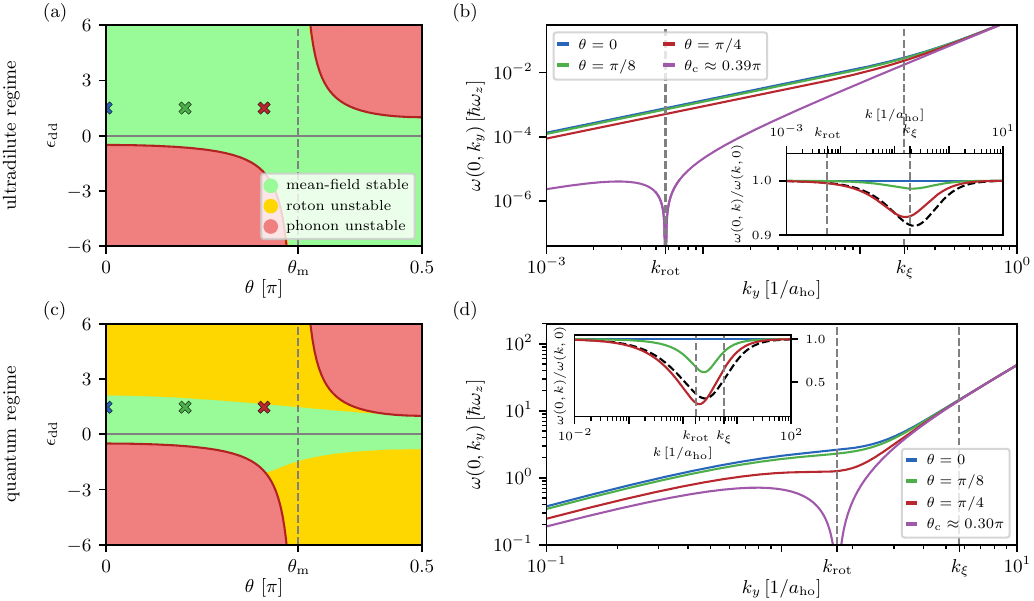}
	\caption{
		Mean-field instability diagram with respect to the dipolar strength $\epsdd$ and the tilting angle $\theta$ in the (a) ultradilute, with $a_\text{s} / \aho \approx 7.5 \cdot 10^{-7}$, and (c)  \quantum case, with $a_\text{s} / \aho \approx 0.019$.
		We show the mean-field stable (green), the roton-unstable (yellow), and the phonon-unstable (red) regions.
		The separation line of the phonon-unstable region is given by $\epsdd (3 \cos^2 \theta - 1) = -1$ and the `magic' angle $\theta_\mathrm{m} = \arccos{(1/\sqrt{3})}\simeq0.3\,\pi$, of vanishing $\epsdd$ dependence is marked by the gray-dashed line.
		For the \udilute case, the instability diagram is dominated by the mean-field stable and the phonon-unstable regime, whereas in the \quantum case much of the stable regime turns roton-unstable.
		Panels (b) and (d) show the Bogoliubov dispersion relations \eqref{eq:Bog_dispersion} along $k_y$ at $k_x=0$ in the \udilute case, for $\epsdd=1.5$, and the \quantum case for $\epsdd=1.47$, for the three tilting angles, as marked by the correspondingly colored crosses in the instability diagrams.
		In addition, we show the dispersion at the phonon (roton) instability, i.e., at $\theta=\theta_\text{c}$ (purple) and mark the corresponding roton momentum $k_\mathrm{rot}$ and the healing momentum $k_\xi$ by gray-dashed lines.
		The insets in (b) and (d) display the ratio $\omega(0,k)/\omega(k,0)$, indicating the region of strongest anisotropy.
		They furthermore show the ratios $\omega_{\epsdd=1.5, \ \theta=0}(k) / \omega_{\epsdd=0, \ \theta=0}(k)$ and $\omega_{\epsdd=1.47, \ \theta=0}(k) / \omega_{\epsdd=0, \ \theta=0}(k)$, respectively, as black dashed lines. 
		These highlight the momentum scales on which the dipolar interaction modifies the dispersion in the isotropic case.
	}
	\label{fig:instab_diag}
\end{figure*}

\Fig{instab_diag} shows the mean-field instability diagram in (a) the \udilute and (c) \quantum cases, depicting the mean-field stable (green), roton-unstable (yellow), and  phonon-unstable (red) regions and their boundaries as determined by the dipolar strength $\epsdd$ and the tilting angle $\theta$.
The phonon-unstable region is determined by the instability of the $k=0$ mode in \eqref{eq:Bog_dispersion} which is equivalent to a negative chemical potential \eqref{eq:chemical_potential_mf} and its critical line is thus parametrized by $\epsdd (3 \cos^2 \theta - 1) = -1$.
The latter is the same in the different parameter regimes since it only depends on the dipolar strength and the tilting angle.
The gray-dashed line indicates the `magic' angle $\theta_\mathrm{m} = \arccos{(1/\sqrt{3})}$, at which the chemical potential is independent of the dipolar interaction and no instability occurs.
Note that for the \quantum parameter choice, stability is significantly reduced due to the increased chemical potential allowing for roton instabilities to occur down to smaller dipolar strengths.

Here we are interested in far-from-equilibrium dynamics, for which we find that even for parameters chosen near but away from the unstable regimes, the system is found to collapse due to local density fluctuations in the Bose gas triggering an instability.
Hence, we chose the dipolar strength and tilting angles as defined in \Subsect{parameters}, sufficiently far away from the instability lines, as marked by the crosses in Figs.~\fig{instab_diag} (a) and (c).

\subsection{Roton vs healing momentum scales}
\label{app:krotheal}

In this appendix, we provide a directional scaling analysis of the momentum spectra.
Panels (b) and (d) of \Fig{instab_diag} show the respective Bogoliubov dispersion relations $\omega(k_{x},k_{y})$, \eqref{eq:Bog_dispersion}, along the $k_y$-axis at $k_x=0$, for $\epsdd=1.5$ (ultradilute) and the \ce{^164 Dy} value $\epsdd=1.47$ (\quantum regime), for different tilting angles, including the critical tilting angle $\theta_\mathrm{c}$ (purple line).
The corresponding roton and healing momenta, $k_\mathrm{rot}$ and $k_\xi=1/\xi_\mathrm{h}$, are marked by gray-dashed lines.
In the \udilute case, one has $k_\mathrm{rot} \simeq 5.7 \cdot 10^{-3} / \aho$, $k_\xi \simeq 0.2 / \aho$, in the \quantum regime $k_\mathrm{rot} \simeq 1.7 / \aho$, $k_\xi \simeq 5.7 / \aho$.

Tilting of the dipoles modifies the isotropy of the dispersion.
The insets in \Fig{instab_diag} (b) and (d) show the ratio $\omega(0,k_{y}=k) / \omega(k_{x}=k,0)$ for the different tilting angles.
While, in the \udilute case, the dispersion is most anisotropic around the healing-length wave number, the anisotropy gets shifted towards the roton wave number in the \quantum case.
The black dashed lines depict the ratios $\omega_{\epsdd, \ \theta=0}(k) / \omega_{\epsdd=0, \ \theta=0}(k)$ with $\epsdd=1.5$ and $1.47$ in the \udilute and \quantum cases, respectively.
In any case, the dispersion is modified in the regime $k< k_\xi$, which is of interest in the context of IR universal coarsening dynamics.

\begin{figure}[t!]
	\centering
	\includegraphics[width=\columnwidth]{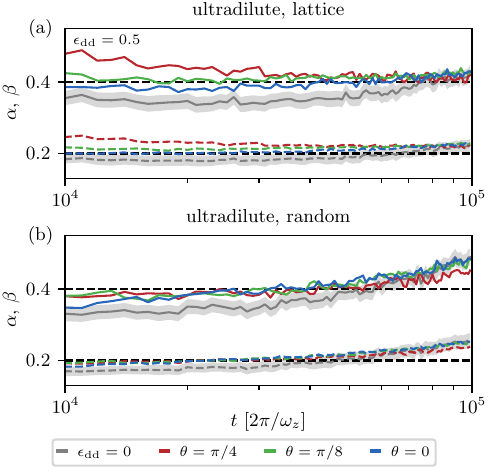}
	\caption{
		Time-varying scaling exponents $\alpha(t)$ (solid) and $\beta(t)$ (dashed) are shown for different tilting angles in the \udilute case at $\epsdd=0.5$ for (a) lattice and (b) random sampling.
		The scaling exponents are obtained as in \Fig{running_rescale_occ_number} and the gray shadow displays the non-dipolar error which is of similar magnitude as the dipolar errors.
	}
	\label{fig:scal_exp_05}
\end{figure}

\section{Scaling exponents at $\epsdd=0.5$}
\label{app:scal_exp_05}

In \Fig{scal_exp_05} we display the scaling exponents $\alpha(t)$ and $\beta(t)$ as functions of time for weaker dipolar strength $\epsdd=0.5$ compared to the exponents shown in \Fig{running_rescale_occ_number} at $\epsdd=1.5$.
All shown exponents are in agreement with the anomalous scaling exponent $\beta\approx1/5$ and fulfill the scaling relation $\alpha=d\beta$.
The universal scaling regime gets significantly increased in comparison with \Fig{running_rescale_occ_number} at strong dipolar interactions which implies a delayed transition to larger scaling exponents at late times.
Due to the delay, the onset of the transition is only observed for random sampling in panel (b) where the onset occurs earlier for weaker tilting angles as discussed in \Subsect{time_dependence_exponents}.
To conclude, we observe and extended scaling regime for weaker dipolar interactions when extracting the exponents using the scaling form \eqref{eq:occupation_number_scaling}.

\section{Anisotropic scaling evolution}
\label{app:anisotropy_spectrum}
%

\begin{figure}[t!]
	\centering
	\includegraphics[width=\columnwidth]{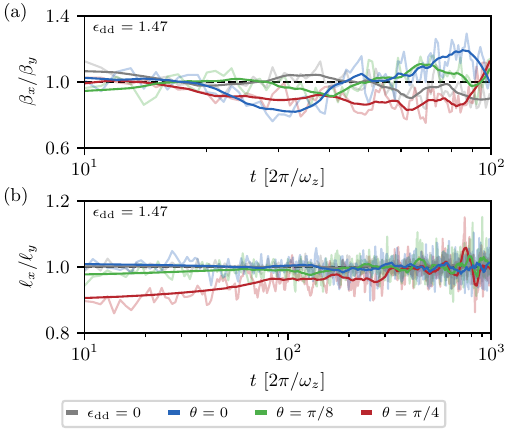}
	\caption{
		Ratios of (a) scaling exponents $\beta_x / \beta_y$ and (b) inter-defect distances $\ell_x / \ell_y$ for the \quantum case, starting from random sampling.
		The scaling exponents are obtained following the scheme outlined in \Fig{running_rescale_occ_number} and matched according to \eqref{eq:occupation_number_axis_averaged_scaling}.
		For panel (b) we separate the average inter-defect distance $\ell$ into its spatial components along the $x$- and $y$-axes.
		The solid curves are obtained by smoothing the raw data shown as shadows using a first-order Savitzky-Golay filter.
		Besides a weak initial anisotropy in the inter-defect distance for the strongly-tilted, dipolar case, no anisotropy is observed throughout the universal scaling regime.	
	}
	\label{fig:anisotropy}
\end{figure}

This appendix discusses the possibility of having anisotropic self-similar evolution and coarsening along the $x$- and $y$- axis. 

In addition to the angle-averaged, single-particle spectra \eqref{eq:single_particle_spectrum}, we also investigated the axis-averaged, single-particle spectra
\begin{align}
	\label{eq:single_particle_spectrum_axis_averaged}
	n_x(k_x,t) &= \int  \dd{k_y} \expval{\psi^*(\vb{k}_\perp, t) \, \psi(\vb{k}_\perp, t)} \,, \notag \\
	n_y(k_y,t) &= \int  \dd{k_x} \expval{\psi^*(\vb{k}_\perp, t) \, \psi(\vb{k}_\perp, t)} \,,
\end{align}
along the $x$- and $y$-axes.
In analogy to \eqref{eq:occupation_number_scaling}, in the vicinity of a non-thermal fixed point, self-similar evolution is expected according to
\begin{align}
	\label{eq:occupation_number_axis_averaged_scaling}
	n_x(k_x,t) &= (t/t_0)^{\alpha_x} n_x((t/t_0)^{\beta_x} k_x, t_0) \,, \notag \\
	n_y(k_y,t) &= (t/t_0)^{\alpha_y} n_y((t/t_0)^{\beta_y} k_y, t_0) \,,
\end{align}
with in general different scaling exponents $\alpha_{x,y}$ and $\beta_{x,y}$.
From particle-number conservation $\alpha_x = d_\mathrm{eff} \beta_x$ and $\alpha_y = d_\mathrm{eff} \beta_y$, with an effective dimensionality $d_\mathrm{eff}$, is implied.
Since one spatial dimension is integrated out in each of the spectra \eqref{eq:single_particle_spectrum_axis_averaged}, the axis-averaged, single-particle spectra are one-dimensional distributions, such that $d_\mathrm{eff}=1$.
In order to extract the scaling exponents $\alpha_x$, $\beta_x$, $\alpha_y$, and $\beta_y$, we performed the same scaling analysis as in Sect.~\ref{sec:self_similar_scaling} over two orders of magnitude in time, taking into account the net imprinted total momentum.

As introduced in \eqref{eq:occupation_number_axis_averaged_scaling}, the anisotropy of the dipolar interaction for non-zero tilting angles poses the question if scaling exponents deviate along the attractive and repulsive direction of the dipoles.
In the \udilute regime, for both initial conditions and all dipolar configurations the equality $\alpha_x\approx\beta_x$ and $\alpha_y\approx\beta_y$ has been confirmed, respectively.
In addition, we found isotropic scaling, with $\beta_x\approx\beta_y$ within errors, throughout.
For most times anomalous scaling $\beta_{x,y} \approx 0.2$ has been found with weakly increasing scaling exponents at late times as in \Fig{running_rescale_occ_number}.
The equality $\beta_x \approx \beta_y$ is also obtained in the \quantum regime, as can be seen in \Fig{anisotropy} (a) for random sampling.

The absence of anisotropic scaling has also been confirmed in coarsening by computing the ratio $\ell_x / \ell_y$ of the average inter-defect distance along the $x$- and $y$-axis, respectively.
In the \udilute regime this ratio is $\approx 1$ for all initial conditions, all dipolar configurations and all times.
In the \quantum case -- for which the results for random sampling are shown in \Fig{anisotropy} (b) -- weak deviations at early times have been observed for non-zero tilting angle, which, however, did not indicate qualitatively different coarsening behavior along different axis.

\section{Clustering in single runs}
\label{app:clustering_single_runs}
%

\begin{figure*}[t]
	\centering
	\includegraphics[width=\textwidth]{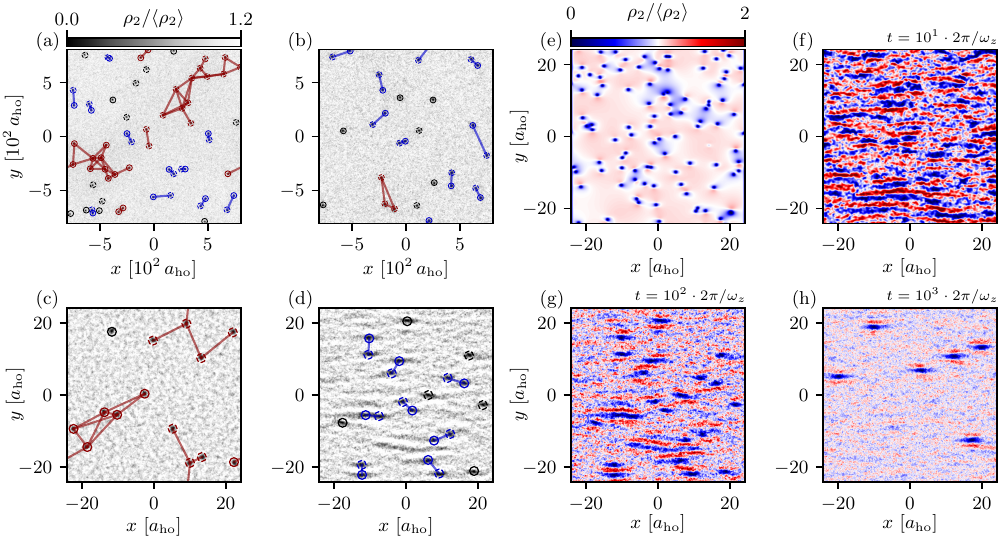}
	\caption{
		Normalized particle density $\rho_2$ for single runs in the \udilute regime at (a) $\epsdd=0$ and (b) $\epsdd=1.5$ and $\theta=\pi/4$ at late times $t=5\cdot10^5  (2\pi)/\omega_z$ starting from lattice sampling.
		In the \quantum regime the dipolar strength $\epsdd=1.47$ at (c) $\theta=0$ and (d) $\theta=\pi/4$ starting from random sampling at late times $t=8\cdot10^2  (2\pi)/\omega_z$ and $t=10^2  (2\pi)/\omega_z$, respectively, is shown.
		Vortices are encircled by a solid and antivortices by a dashed line and afterwards classified into isolated vortices (black), vortex dipoles (blue) and vortex clusters (red) using the algorithm described below.
		The connection lines between vortex dipoles and vortex clusters specify to which dipole or cluster the respective vortex belongs.
		Panels (e-h) show the normalized particle density in the \quantum regime for $\epsdd=1.47$ and strong tilting $\theta=\pi/4$ starting from random sampling for various times.
	}
	\label{fig:clustering_single_run}
\end{figure*}

In this section the observed (anti)clustering behavior based on the degree of clustering $P_\mathrm{c}$, cf.~\Fig{clustering}, and the ratio of the equal-sign and opposite-sign inter-defect distance, cf.~\Fig{clustering}, is showcased in single realizations.
For this we classify vortices into three categories -- clusters, dipoles, isolated vortices -- based on an algorithm put forward in \cite{Reeves2013a.PhysRevLett.110.104501,Valani2018a} and briefly outlined in the following.

We start by picking a vortex $i$ and its nearest-neighboring opposite-sign vortex $j$ out of the defect configuration.
Within the circle spanned around vortex $i$, with radius being the distance between defects $i$ and $j$, we mark all equal-sign vortices, if existent, as possible cluster candidates of vortex $i$.
If none are found, the defect $j$ is marked as possible dipole candidate of defect $i$. 
This procedure is repeated for every defect such that every directed link between two defects is labeled as either `possibly clustered', `possibly dipole' or `no connection'.
If the mutual links between two defects have both been labeled as `possibly clustered' or `possibly dipole', then the undirected link is labeled `clustered' or `dipole', respectively.
All defects without a `clustered' or `dipole' link are labeled `isolated'.
By this we obtain a connection matrix containing all links between the vortex dipoles and clusters from which the individual clusters and dipoles can be extracted by determining all connected components.

In \Fig{clustering_single_run} this classification is displayed together with all connection lines in the \udilute regime for (a) the non-dipolar $\epsdd=0$ and (b) the strongly-tilted, dipolar $\epsdd=1.5$ case at late times, $t=5\cdot10^5 (2 \pi)/\omega_z$, starting from lattice sampling.
The solid rings indicate elementary vortices with $q=1$, whereas dashed rings indicate antivortices with $q=-1$, and the color code distinguishes between clustered (red), dipole (blue), and isolated (black) defects.
From the connection lines it becomes apparent, that the above classification scheme does not necessarily imply every vortex within a cluster to be connected with any other vortex of the same cluster.
In the non-dipolar case (a), the emergence of two large vortex clusters, which extend over the periodic boundary, with vortex dipoles in between, are observed, in agreement with the observed clustering in \Fig{clustering} (a,d).
For the strongly-tilted, dipolar case (b) the late time vortex configuration consists predominantly of dipoles, as has been found in \Fig{clustering} (a,d).

\Fig{clustering_single_run} further shows the defect classification in the \quantum regime at dipolar strength $\epsdd=1.47$ in (c) the isotropic and (d) the strongly tilted case, at late times $t=8\cdot10^2 (2\pi)/\omega_z$ and $t=10^2  (2\pi)/\omega_z$, respectively.
The isotropic case exhibits the formation of vortex clusters whereas the strongly-tilted case is dominated by isolated vortices and vortex dipoles, agreeing with \Fig{clustering} (c,f).
Thus, the (anti)clustering behavior discussed in Sects.~\ref{sec:clustering} can already be found in single realizations.

In panels (e-h) in \Fig{clustering_single_run} an exemplary time evolution is shown for strong tilting $\theta=\pi/4$ at $\epsdd=1.47$ in the \quantum regime starting from random sampling.
The early dynamics appears to be dominated by horizontal, stripe-shaped density fluctuations on the order of the roton momentum $k_\mathrm{rot}$, cf.~\Fig{instab_diag}, indicating the presence of rotonic excitations.
At later times the presence of roton-like excitations turns localized around the vortex defects in the form of density ripples.
The possible implications of this altered vortex shape on vortex-antivortex annihilation processes and subsequently on the clustering behavior of the system have been discussed in the main text, cf.~\Subsect{degree_of_clustering}.

\end{appendix}
\twocolumngrid

\bibliographystyle{apsrev4-2}

\providecommand*\hyphen{-}
%

\end{document}